\documentclass[ALICE,manyauthors]{cernphprep}
\usepackage[comma,square,numbers,sort&compress]{natbib}
\usepackage{hyperref}
\usepackage{lineno}
\usepackage{xspace}
\usepackage{tabu}
\usepackage{multirow}
\usepackage{soul}
\usepackage{rotating, graphicx}
\usepackage{lscape}
\usepackage[T1]{fontenc}
\usepackage{float}

\usepackage{color}
\definecolor{orange}{rgb}{1,0.5,0}
\definecolor{darkgreen}{rgb}{0.0, 0.6, 0.22}

\begin{document}

%%%%%%%%%%%%%%%%%%%%%%%%%%%%%%%%%%%%%%%%%%%%%%%%%%
% These are some new commands that may be useful 
% for paper writing in general. If other newcommands
% are needed for your specific paper, please feel 
% free to add here. 
%
% The currently available commands are organized in: 
% 1) Systems
% 2) Quantities
% 3) Energies and units
% 4) Detectors
% 5) particle species 
%%%%%%%%%%%%%%%%%%%%%%%%%%%%%%%%%%%%%%%%%%%%%%%%%%

% 1) SYSTEMS 
\newcommand{\pp}           {pp\xspace}
\newcommand{\ppbar}        {\mbox{$\mathrm {p\overline{p}}$}\xspace}
\newcommand{\XeXe}         {\mbox{Xe--Xe}\xspace}
\newcommand{\PbPb}         {\mbox{Pb--Pb}\xspace}
\newcommand{\pA}           {\mbox{pA}\xspace}
\newcommand{\pPb}          {\mbox{p--Pb}\xspace}
\newcommand{\AuAu}         {\mbox{Au--Au}\xspace}
\newcommand{\dAu}          {\mbox{d--Au}\xspace}

% 2) QUANTITIES 
\newcommand{\s}            {\ensuremath{\sqrt{s}}\xspace}
\newcommand{\snn}          {\ensuremath{\sqrt{s_{\mathrm{NN}}}}\xspace}
\newcommand{\pt}           {\ensuremath{p_{\rm T}}\xspace}
\newcommand{\meanpt}       {$\langle p_{\mathrm{T}}\rangle$\xspace}
\newcommand{\ycms}         {\ensuremath{y_{\rm CMS}}\xspace}
\newcommand{\ylab}         {\ensuremath{y_{\rm lab}}\xspace}
\newcommand{\etarange}[1]  {\mbox{$\left | \eta \right |~<~#1$}}
\newcommand{\yrange}[1]    {\mbox{$\left | y \right |<0.5$}}
\newcommand{\dndy}         {\ensuremath{\mathrm{d}N_\mathrm{ch}/\mathrm{d}y}\xspace}
\newcommand{\dndeta}       {\ensuremath{\mathrm{d}N_\mathrm{ch}/\mathrm{d}\eta}\xspace}
\newcommand{\avdndeta}     {\ensuremath{\langle\dndeta\rangle}\xspace}
\newcommand{\dNdy}         {\ensuremath{\mathrm{d}N_\mathrm{ch}/\mathrm{d}y}\xspace}
\newcommand{\Npart}        {\ensuremath{N_\mathrm{part}}\xspace}
\newcommand{\Ncoll}        {\ensuremath{N_\mathrm{coll}}\xspace}
\newcommand{\dEdx}         {\ensuremath{\textrm{d}E/\textrm{d}x}\xspace}
\newcommand{\RpPb}         {\ensuremath{R_{\rm pPb}}\xspace}
\newcommand{\vn}{\ensuremath{v_{\rm n}}}
\newcommand{\vnpt}{\ensuremath{v_{\rm n}(p_{\mathrm{T}})}}
\newcommand{\vtwopt}{\ensuremath{v_{\rm 2}(p_{\mathrm{T}})}}
\newcommand{\vthreept}{\ensuremath{v_{\rm 3}(p_{\mathrm{T}})}}
\newcommand{\vtwomax}{\ensuremath{\pt{}\vert_{v_2^{\mathrm{max}}}}}

% 3) ENERGIES, UNITS
\newcommand{\nineH}        {$\sqrt{s}~=~0.9$~Te\kern-.1emV\xspace}
\newcommand{\seven}        {$\sqrt{s}~=~7$~Te\kern-.1emV\xspace}
\newcommand{\twoH}         {$\sqrt{s}~=~0.2$~Te\kern-.1emV\xspace}
\newcommand{\twosevensix}  {$\sqrt{s}~=~2.76$~Te\kern-.1emV\xspace}
\newcommand{\five}         {$\sqrt{s}~=~5.02$~Te\kern-.1emV\xspace}
\newcommand{\twosevensixnn}{$\sqrt{s_{\mathrm{NN}}}~=~2.76$~Te\kern-.1emV\xspace}
\newcommand{\fivenn}       {$\sqrt{s_{\mathrm{NN}}}~=~5.02$~Te\kern-.1emV\xspace}
\newcommand{\LT}           {L{\'e}vy-Tsallis\xspace}
\newcommand{\GeVc}         {Ge\kern-.1emV/$c$\xspace}
\newcommand{\MeVc}         {Me\kern-.1emV/$c$\xspace}
\newcommand{\TeV}          {Te\kern-.1emV\xspace}
\newcommand{\GeV}          {Ge\kern-.1emV\xspace}
\newcommand{\MeV}          {Me\kern-.1emV\xspace}
\newcommand{\GeVmass}      {Ge\kern-.1emV/$c^2$\xspace}
\newcommand{\MeVmass}      {Me\kern-.1emV/$c^2$\xspace}
\newcommand{\lumi}         {\ensuremath{\mathcal{L}}\xspace}

% 4) DETECTORS 
\newcommand{\ITS}          {\rm{ITS}\xspace}
\newcommand{\TOF}          {\rm{TOF}\xspace}
\newcommand{\ZDC}          {\rm{ZDC}\xspace}
\newcommand{\ZDCs}         {\rm{ZDCs}\xspace}
\newcommand{\ZNA}          {\rm{ZNA}\xspace}
\newcommand{\ZNC}          {\rm{ZNC}\xspace}
\newcommand{\SPD}          {\rm{SPD}\xspace}
\newcommand{\SDD}          {\rm{SDD}\xspace}
\newcommand{\SSD}          {\rm{SSD}\xspace}
\newcommand{\TPC}          {\rm{TPC}\xspace}
\newcommand{\TRD}          {\rm{TRD}\xspace}
\newcommand{\VZERO}        {\rm{V0}\xspace}
\newcommand{\VZEROA}       {\rm{V0A}\xspace}
\newcommand{\VZEROC}       {\rm{V0C}\xspace}
\newcommand{\Vdecay} 	   {\ensuremath{V^{0}}\xspace}

% 4) PARTICLE SPECIES 
\newcommand{\ee}           {\ensuremath{e^{+}e^{-}}} 
\newcommand{\pip}          {\ensuremath{\pi^{+}}\xspace}
\newcommand{\pim}          {\ensuremath{\pi^{-}}\xspace}
\newcommand{\kap}          {\ensuremath{\rm{K}^{+}}\xspace}
\newcommand{\kam}          {\ensuremath{\rm{K}^{-}}\xspace}
\newcommand{\pbar}         {\ensuremath{\rm\overline{p}}\xspace}
\newcommand{\kzero}        {\ensuremath{{\rm K}^{0}_{\rm{S}}}\xspace}
\newcommand{\lmb}          {\ensuremath{\Lambda}\xspace}
\newcommand{\almb}         {\ensuremath{\overline{\Lambda}}\xspace}
\newcommand{\Om}           {\ensuremath{\Omega^-}\xspace}
\newcommand{\Mo}           {\ensuremath{\overline{\Omega}^+}\xspace}
\newcommand{\X}            {\ensuremath{\Xi^-}\xspace}
\newcommand{\Ix}           {\ensuremath{\overline{\Xi}^+}\xspace}
\newcommand{\Xis}          {\ensuremath{\Xi^{\pm}}\xspace}
\newcommand{\Oms}          {\ensuremath{\Omega^{\pm}}\xspace}
\newcommand{\degree}       {\ensuremath{^{\rm o}}\xspace}
\newcommand{\pipm}         {\ensuremath{\pi^{\pm}}\xspace}
\newcommand{\kapm}          {\ensuremath{\rm{K}^{\pm}}\xspace}
\newcommand{\lambdas}{$\Lambda$+$\overline{\Lambda}$}
\newcommand{\vo}{\ensuremath{{\rm V}^{\rm 0}}}

% 5) Help 
\newcommand{\todo}[1]{\textcolor{darkgreen}{[TODO: \emph{#1}]}}
\newcommand{\done}[1]{\textcolor{blue}{[DONE: \emph{#1}]}}
\newcommand{\new}[1]{\textcolor{red}{#1}}
\newcommand{\change}[1]{\textcolor{blue}{[CHANGE: #1]}}

%%%%%%%%%%%%%%%  Title page %%%%%%%%%%%%%%%%%%%%%%%%
\begin{titlepage}
% the dates below correspond to CERN approval
% please don't touch: EB chairs will take care
\PHyear{2021}       % required, will be obtained from CERN
\PHnumber{149}      % required, will be obtained from CERN
\PHdate{21 July}  % required, will be obtained from CERN
%%%%%%%%%%%%%%%%%%%%%%%%%%%%%%%%%%%%%%%%%%%%%%%%%%%%

%%% Put your own title + short title here:
\title{Anisotropic flow of identified hadrons in \XeXe collisions at $\snn = 5.44$~TeV}
\ShortTitle{PID flow in \XeXe{} collisions}   % appears on left page headers

%%% Do not change the next lines
\Collaboration{ALICE Collaboration\thanks{See Appendix~\ref{app:collab} for the list of collaboration members}}
\ShortAuthor{ALICE Collaboration} % appears on right page headers, do not change

\begin{abstract}

Measurements of elliptic ($v_2$) and triangular ($v_3$) flow coefficients of \pipm{}, \kapm{}, p+\pbar{}, \kzero{}, and \lambdas{} obtained with the scalar product method in \XeXe collisions at $\snn = 5.44$~TeV are presented. The results are obtained in the rapidity range \yrange{} and reported as a function of transverse momentum, \pt{}, for several collision centrality classes. The flow coefficients exhibit a particle mass dependence for $\pt<3$~\GeVc, while a grouping according to particle type (i.e., meson and baryon) is found at intermediate transverse momenta ($3<\pt<8$~\GeVc). The magnitude of the baryon $v_{2}$ is larger than that of mesons up to $\pt=6$~\GeVc. The centrality dependence of the shape evolution of the \pt-differential $v_2$ is studied for the various hadron species. The $v_2$ coefficients of \pipm{}, \kapm{}, and p+\pbar{} are reproduced by MUSIC hydrodynamic calculations coupled to a hadronic cascade model (UrQMD) for $\pt<1$~\GeVc. A comparison with $v_{\rm n}$ measurements in the corresponding centrality intervals in Pb--Pb collisions at $\snn = 5.02$~TeV yields an enhanced $v_2$ in central collisions and diminished value in semicentral collisions.

\end{abstract}
\end{titlepage}

\setcounter{page}{2} %please do not remove this line

%%%%%%%%%%%%%%%%%%%%%%%%%%%%%%%%
% begin main text
%%%%%%%%%%%%%%%%%%%%%%%%%%%%%%%%

\section{Introduction}
\label{sec:intro}

Collisions of ultra-relativistic nuclei provide the opportunity to study in the laboratory the quark--gluon plasma (QGP), a state of deconfined quarks and gluons~\cite{Bass:1998vz}. An important feature of the QGP is the collective expansion, called flow, due to pressure gradients in the geometrically overlapping matter in the collisions of nuclei. A direct experimental evidence of this collective flow is the observation of anisotropic flow~\cite{Ollitrault:1992bk}, which arises from the asymmetry in the initial geometry of the collision combined with the initial state inhomogeneities of the system's energy density. Its magnitude is usually quantified by the harmonic coefficients $v_{\rm n}$ in a Fourier decomposition of the azimuthal distribution of particles with respect to the collision symmetry plane~\cite{Voloshin:1994mz, Poskanzer:1998yz}

\begin{equation}
\label{eq:flowdud}
\frac{{\rm d}N}{{\rm d}\varphi} \propto 1+2\sum_{{\rm n}=1}^{\infty}v_{\rm n} \cos[{\rm n}(\varphi - \Psi_{\rm n})],
\end{equation}

where $\varphi$ is the azimuthal angle of the produced particle and $\Psi_{\rm n}$ is the $n$-th harmonic symmetry-plane angle in the collision. The second ($v_2$) and third ($v_3$) coefficients are called elliptic and triangular flow, respectively. While $v_2$ directly reflects the almond-shaped geometry of the interaction volume being the largest contribution to the asymmetry in non-central collisions, $v_3$ is generated by fluctuations in the initial distribution of nucleons in the overlap region \cite{Bhalerao:2006tp, Alver:2008zza, Takahashi:2009na, Alver:2010gr, Alver:2010dn}. For light and strange particles, both coefficients scale approximately linearly with the corresponding eccentricities $\varepsilon_{\rm n}$ ($v_{\rm n} \approx \kappa_{\rm n}\varepsilon_{\rm n}$)~\cite{Gardim:2011xv}, which govern the shape of the initial collision geometry. The coefficients $\kappa_{\rm n}$ are sensitive to the macroscopic properties of the QGP, such as the shear viscosity to entropy density ratio ($\eta/s$), and the lifetime of the system. A greater sensitivity to $\eta/s$ is expected for higher-order flow coefficients~\cite{Qin:2010pf, Teaney:2010vd}.

Measurements of anisotropic flow performed in Au--Au collisions at the Relativistic Heavy Ion Collider (RHIC)~\cite{Arsene:2004fa, Adcox:2004mh, Back:2004je, Adams:2005dq} and in Pb--Pb collisions at the Large Hadron Collider (LHC)~\cite{ALICE:2011ab, Acharya:2018lmh, ATLAS:2012at, Chatrchyan:2013kba} indicate that the QGP is strongly-coupled (i.e. constituents have small mean free path) and behaves like a nearly perfect fluid as the extracted $\eta/s$ is close to the lower limit predicted by the anti-de Sitter/conformal field theory (AdS/CFT) correspondence of $1/(4\pi)$ (setting $\hslash=k_{\rm B}=1$)~\cite{Kovtun:2004de}. Recently, the $v_{\rm n}$ coefficients of unidentified charged particles have been measured in Xe--Xe collisions at the center-of-mass energy per nucleon pair $\snn = 5.44$~TeV~\cite{Acharya:2018ihu, Sirunyan:2019wqp, Aad:2019xmh}. These measurements further constrain the transport coefficients of the medium, such as $\eta/s$ and bulk viscosity to entropy density ratio ($\zeta/s$), and initial state models. Furthermore, comparisons of the $v_2$ measurements in semicentral Xe--Xe collisions with those from Pb--Pb collisions in the same centrality intervals could provide direct information on the $\eta/s$. For these collisions, the two systems have similar $\varepsilon_{2}$ coefficients~\cite{Giacalone:2017dud, Eskola:2017bup} but different sizes, thus the influence of the initial state on $\eta/s$ mostly cancels out in ratios of Xe--Xe/Pb--Pb $v_2$ and a finite $\eta/s$ suppresses $\kappa_2$ by 1/$R$, where $R$ corresponds to the transverse size of the system~\cite{Giacalone:2017dud}. Centrality estimates the degree of overlap between two colliding nuclei and is expressed as percentiles of the inelastic cross section, with low percentage values corresponding to the most central collisions. Stronger constraints can be placed by studying anisotropic flow of identified particles since the $\kappa_{\rm n}$ coefficients depend on particle mass, type, and kinematics~\cite{Qiu:2011iv}. In addition to probing $\eta/s$ and $\zeta/s$, the anisotropic flow of identified particles provides valuable information on the particle production mechanism in different transverse momentum, \pt{}, regions. For ${\pt \lesssim}$~3~\GeVc{}, the characteristic mass ordering (i.e., lighter particles having a larger $v_{\rm n}$ than that of heavier particles at fixed \pt{}), which arises from the interplay between radial flow (isotropic expansion) and anisotropic flow~\cite{Huovinen:2001cy, Shen:2011eg}, is described by hydrodynamic calculations~\cite{Abelev:2014pua, Adam:2016nfo, Acharya:2018zuq, Adams:2003am, Adler:2003kt}. This mass ordering provides constraints on both $\eta/s$ and $\zeta/s$ as the magnitude of $v_{\rm n}$ depends on $\eta/s$, while the mass ordering is affected by $\zeta/s$ through its influence on radial flow. At intermediate \pt{}, ${3 < \pt < 8}$~\GeVc{}, a grouping of $v_{\rm n}$ of mesons and baryons is observed, with the flow of baryons being larger than that of mesons~\cite{Acharya:2018zuq, Abelev:2012di, Abelev:2007qg, Adare:2012vq}. While this supports the hypothesis of hadronization through quark coalescence (involving the combination of a quark and anti-quark to form a meson and three quarks to form a baryon)~\cite{Molnar:2003ff, Greco:2003mm, Fries:2003kq}, alternate explanations are attempted in models in which particle production includes interactions of jet fragments with bulk matter~\cite{Werner:2012xh}. To test the hypothesis of particle production via quark coalescence it was suggested to divide both $v_{\rm n}$ and \pt{} by the number of constituent quarks since it is assumed that the spectrum of produced particles is proportional to the product of the spectra of their constituents~\cite{Sato:1981ez, Dover:1991zn}. However, deviations from the exact scaling at the level of $\pm 20\%$ are seen in Pb--Pb collisions at the LHC~\cite{Abelev:2014pua, Adam:2016nfo, Acharya:2018zuq}, while it only holds approximately at RHIC \cite{Adare:2012vq}. This scaling can be further tested using measurements of identified particle $v_{\rm n}$ in Xe--Xe collisions.

The \pt-differential elliptic flow coefficient, \vtwopt{}, of \pipm{}, \kapm{}, p+\pbar{}, \kzero{}, and \lambdas{} as well as the \pt-differential triangular flow coefficient, \vthreept{}, of \pipm{}, \kapm{}, and p+\pbar{}, measured in Xe--Xe collisions at $\sqrt{s_{\rm NN}}$~=~5.44 TeV are presented in this paper. The results are reported for $\pt < 8.5$ \GeVc{} within the rapidity range $\vert y \vert <$ 0.5 at different collision centralities in the 0--60\% range, where $v_{\rm n}$ can be measured accurately. The scalar product method~\cite{Adler:2002pu, Voloshin:2008dg, Luzum:2012da} is employed with a pseudorapidity gap of $|\Delta\eta|>2.0$ between the identified particles under study and the reference charged particles. The $v_{\rm n}$ coefficients denote the average between results for positive and negative particles as they are compatible within uncertainties for most \pt{} and centrality intervals. Any residual difference has been included into the systematic uncertainties.

This paper is organized as follows. A brief description of the ALICE detector, analysis details, particle identification, reconstruction methods, and flow measurement techniques is given in Sec.~\ref{sec:analysis}. Section~\ref{sec:systematics} outlines the evaluation of systematic uncertainties, while the results are reported in Sec.~\ref{sec:results}. Finally, conclusions are drawn in Sec.~\ref{sec:summary}.

\section{Experimental setup and analysis details}
\label{sec:analysis}

A full overview of the ALICE detector and its performance can be found in Refs.~\cite{Aamodt:2008zz, Abelev:2014ffa}. The Inner Tracking System (ITS)~\cite{Aamodt:2010aa}, the Time Projection Chamber (TPC)~\cite{Alme:2010ke}, the Time of Flight (TOF)~\cite{Akindinov:2013tea}, and the V0~\cite{Abbas:2013taa} are the main subsystems used in this analysis and are briefly described below. These detectors are located inside a solenoid magnet which provides a nominal magnetic field of 0.5 T. However, the field was reduced to 0.2 T for Xe--Xe collisions in order to extend particle tracking and identification to the lowest possible momenta. The ITS, TPC, and TOF detectors cover the full azimuth within the pseudorapidity range $|\eta|<0.9$. The ITS consists of six layers of silicon detectors and is employed for tracking, vertex reconstruction, and event selection. The TPC, being the main tracking detector, is used to reconstruct charged-particle tracks but also to identify particles via the measurement of the specific energy loss, \dEdx{}. The TOF detector provides particle identification based on the measurement of flight time from the collision point using a start time given by the T0 detector~\cite{Bondila:2005xy}, which consists of two arrays of Cherenkov counters located at ${-3.3 < \eta < -3.0}$~(T0C) and ${4.5 < \eta < 4.9}$~(T0A). The V0 detector, two arrays of 32 scintillator tiles each (four rings in the radial direction with each ring divided into eight sectors in the azimuthal direction) covering ${-3.7 < \eta < -1.7}$~(V0C) and ${2.8 < \eta < 5.1}$~(V0A), is used for triggering, event selection, and the determination of centrality~\cite{ALICE-PUBLIC-2018-003} and ${\bf Q}_{\rm n}$ vectors (see below). Two tungsten-quartz neutron Zero Degree Calorimeters (ZDCs)~\cite{Arnaldi:1999zz}, installed 112.5 meters from the interaction point on each side, are also used for event selection.

The analyzed data set was recorded by the ALICE detector during the Xe--Xe run at $\sqrt{s_{\rm NN}} = 5.44$~TeV in 2017. The minimum-bias trigger requires signals in both V0A and V0C detectors in coincidence with signals in the two neutron ZDCs, the latter condition suppressing contamination from electromagnetic interactions. In addition, the beam-induced background (i.e., beam--gas events) and pileup events are removed using an offline event selection. The former is rejected utilizing the V0 and ZDC timing information, while pileup events are removed by comparing charged particle multiplicity estimates from the V0 detector with those of tracking detectors at midrapidity, exploiting the difference in readout times between the systems. The remaining contribution of such interactions is estimated to be negligible. The primary vertex position is determined from tracks reconstructed in the ITS and TPC as described in Ref.~\cite{Abelev:2014ffa}. Approximately $9\times 10^5$ Xe--Xe events in the 0--60\% centrality interval, with a primary vertex position within $\pm 10$~cm from the nominal interaction point along the beam direction, are used in the analysis. Centrality is estimated from the energy deposition measured in the V0 detector~\cite{ALICE-PUBLIC-2018-003}.

The charged particle tracks used to determine the flow coefficients of \pipm, \kapm, and p+\pbar are reconstructed using the ITS and TPC within ${|\eta|<0.8}$ and ${0.4 < \pt < 8.5}$~\GeVc{}. Each track is required to cross at least 70 TPC readout rows (out of a maximum of 159), to have a minimum number of 70 TPC space points with a $\chi^2$ per TPC space point lower than 4, and to have the ratio between the number of space points and the number of crossed rows in the TPC larger than 0.8. The selected tracks are also required to have at least 2 ITS hits, of which at least one in the two innermost layers, and a $\chi^2$ per ITS hit smaller than 36. Only tracks with a distance of closest approach (DCA) to the reconstructed vertex position smaller than 2 cm in the longitudinal direction ($z$) are accepted. In the transverse plane ($xy$), a \pt-dependent selection is applied: $|{\rm DCA}_{xy}|<7\sigma_{{\rm DCA}_{xy}}$, where $\sigma_{{\rm DCA}_{xy}}$ is the resolution of the ${\rm DCA}_{xy}$ in each \pt{} interval. These selection criteria reduce the contamination from secondary charged particles (i.e., particles originating from weak decays, conversions, and secondary hadronic interactions in the detector material) and fake tracks (random associations of space points) and ensure a track momentum resolution better than 4\% for the considered \pt{} range~\cite{Acharya:2018eaq}. 

The particle identification for \pipm, \kapm, and p+\pbar{} is performed using signals from the TPC and TOF detectors following the procedure described in Ref.~\cite{Acharya:2018zuq}. For $\pt<4$ \GeVc{}, particle identification is done track-by-track evaluating the difference between the measured and expected \dEdx{} and time-of-flight for a given species in units of the standard deviation ($\sigma_{\rm TPC}, \sigma_{\rm TOF}$) from the most probable value. Particles are selected combining the TPC and TOF information (${\rm n}_{\sigma_{\rm PID}} = \sqrt{{\rm n}^{2}_{\sigma_{\rm TPC}} + {\rm n}^{2}_{\sigma_{\rm TOF}}}$) and requiring ${\rm n}_{\sigma_{\rm PID}} < 3$ for each species. When this condition is fulfilled by more than one species, the smallest ${\rm n}_{\sigma_{\rm PID}}$ is used to assign the identity. To exclude contamination in the sample from secondary protons originating from the detector material, only \pbar{} are considered for $\pt < 2$ \GeVc{}. For $\pt > 4$~\GeVc{}, only \pipm{} and p+\pbar{} are identified using the TPC \dEdx{} by selecting them from the upper part of the pion \dEdx{} distribution and from the lower part of the proton \dEdx{} distribution, respectively. For example, pion selection varies in the range $0.3\sigma$ to $2\sigma$. 

The remaining contamination from secondary particles originating in weak decays, studied using the procedure described in Ref.~\cite{Abelev:2013vea}, is negligible for \kapm{} and decreases with increasing \pt from about 5\% to 0.5\% for \pipm{} and from about 40\% to 5\% for p+\pbar{} in the \pt{} range 0.4--4.0~\GeVc{}. The $v_{\rm n}$ coefficients are not corrected for these contaminations. Their effect on $v_{\rm n}$, estimated from the correlation between $v_{\rm n}$ and contamination for various ${\rm DCA}_{xy}$ selections in each \pt{} interval, is negligible for \pipm{} and \kapm{} and up to 20\% and 5\% for central and peripheral collisions, respectively, for $v_2$ of p+\pbar{} at $\pt \sim 0.5$~\GeVc{}. The contamination from other particle species is below 2\% and 25\% at \pt{} $>$ 4.0 \GeVc{} for \pipm{} and p+\pbar{}, respectively. The contamination from fake tracks is negligible. 

The \kzero{} and \lambdas{} are reconstructed in the \kzero{} $\rightarrow$ \pip{} + \pim{} and $\Lambda{}$ $\rightarrow$ p + \pim{} ($\overline{\Lambda}$ $\rightarrow$ \pbar{} + \pip{}) channels. An offline selection is used to identify secondary vertices (called V$^0$s), from which two particles of opposite charge originate. The selection of V$^0$ candidates is done with an invariant mass between 0.4 and 0.6 \GeVmass{} for \kzero{} and 1.07 and 1.17 \GeVmass{} for \lambdas{}. Daughter particles, identified using the TPC ($\vert {\rm n}_{\sigma_{\rm TPC}} \vert < 3$), are assumed to be either a \pip--\pim{} pair or a p--\pim{} (\pbar--\pip{}) pair in the calculation of the invariant mass of the \vo{}. The TPC track quality requirements described above for charged tracks are also imposed on daughter particles. In addition, the maximum DCA of daughter tracks to the secondary vertex is 0.5~cm and the minimum DCA of daughter tracks to the primary vertex is 0.1~cm. Secondary vertices created by decays into more than two particles are rejected requiring the cosine of the pointing angle $\theta_{\rm p}$ to be larger than 0.998.  This angle is defined as the angle between the momentum-vector of the \vo{} assessed at its decay position and the line connecting the \vo{} decay vertex to the primary vertex and has to be close to 0 as a result of momentum conservation. Only \vo{} candidates produced at a radial distance between 5 and 100~cm from the beam line are accepted. Finally, a selection in the Armenteros--Podolanski variables~\cite{armpod} is applied for the \kzero{} candidates to asses the systematic uncertainty related to contamination from \lambdas{} and electron--positron pairs coming from $\gamma$ conversions. Earlier studies have shown that contaminations from higher mass baryons (\Xis{}, \Oms{}) have a negligible effect on the measured $v_{\rm n}$~\cite{Abelev:2014pua}. More details about this selection can be found in Ref.~\cite{Acharya:2018zuq}.

The scalar product (SP) method~\cite{Adler:2002pu, Voloshin:2008dg, Luzum:2012da} is used to measure the flow coefficients $v_{\rm n}$, written as 
\begin{equation}
    v_{\rm n}\{{\rm SP}\} = \langle \langle {\bf u}_{\rm n, k} {\bf Q}_{\rm n}^{*} \rangle \rangle \Bigg/ \sqrt{ \frac{\langle {\bf Q}_{\rm n}  {\bf Q}_{\rm n}^{\rm A *} \rangle \langle  {\bf Q}_{\rm n} {\bf Q}_{\rm n}^{\rm B *} \rangle} { \langle  {\bf Q}_{\rm n}^{\rm A} {\bf Q}_{\rm n}^{\rm B *} \rangle } },
\label{eq:mth_sp}
\end{equation}
where ${\bf u}_{\rm n, k} =\exp(i{\rm n}\varphi_{\rm k})$ is the unit flow vector of the particle of interest $k$ with azimuthal angle $\varphi_{\rm k}$, ${\bf Q}_{\rm n}$ is the event flow vector, and $n$ is the harmonic number. Brackets $\langle \cdots \rangle$ denote an average over all events, the double brackets $\langle \langle \cdots \rangle \rangle$ an average over all particles in all events, and $^*$ the complex conjugate. The vector ${\bf Q}_{\rm n}$ is obtained from the azimuthal distribution of the energy deposition measured in the V0A, with the $x$ and $y$ components given by
\begin{equation}
\label{eq:Qcomp}  
Q_{\rm n,x} = \sum_{\rm j} w_{\rm j} \cos({\rm n} \varphi_{\rm j}), \; Q_{\rm n,y} = \sum_{\rm j} w_{\rm j} \sin({\rm n} \varphi_{\rm j}),
\end{equation}
where the sum runs over the 32 channels $j$ of the V0A detector, $\varphi_{\rm j}$ is the azimuthal angle of channel $j$, and $w_{\rm j}$ is the amplitude measured in channel $j$. The vectors ${\bf Q}_{\rm n}^{\rm A}$ and ${\bf Q}_{\rm n}^{\rm B}$ are determined from the azimuthal distribution of the energy deposition measured in the V0C and the azimuthal distribution of the tracks reconstructed in the ITS and TPC, respectively. Any non-uniform detector response is taken into account by adjusting the components of the ${\bf Q}_{\rm n}$ vectors using a recentering procedure (i.e.\ subtraction of the ${\bf Q}_{\rm n}$ vector averaged over many events from the ${\bf Q}_{\rm n}$ vector of each event)~\cite{Selyuzhenkov:2007zi}. The large gap in pseudorapidity between \textbf{u}$_{\rm n, k}$ and ${\bf Q}_{\rm n}$ ($|\Delta\eta|>2.0$) greatly suppresses short-range correlations unrelated to the common symmetry planes $\Psi_{\rm n}$ (``non-flow"), such as those due to resonances, jets, and quantum statistics correlations. The remaining non-flow contributions are small as reported in Ref.~\cite{Acharya:2018ihu} where the ratio between $v_2\{4\}$ and $v_2\{2, |\Delta\eta|>2.0\}$ of inclusive charged particles shows a weak centrality dependence for semicentral and peripheral collisions. These contributions are estimated by increasing the pseudorapidity gap to $|\Delta\eta|>2.8$. Any residual difference has been included into the systematic uncertainties (see Sec.~\ref{sec:systematics}).
\begin{figure}[!t]
    \begin{center}
    \includegraphics[width=.7\textwidth]{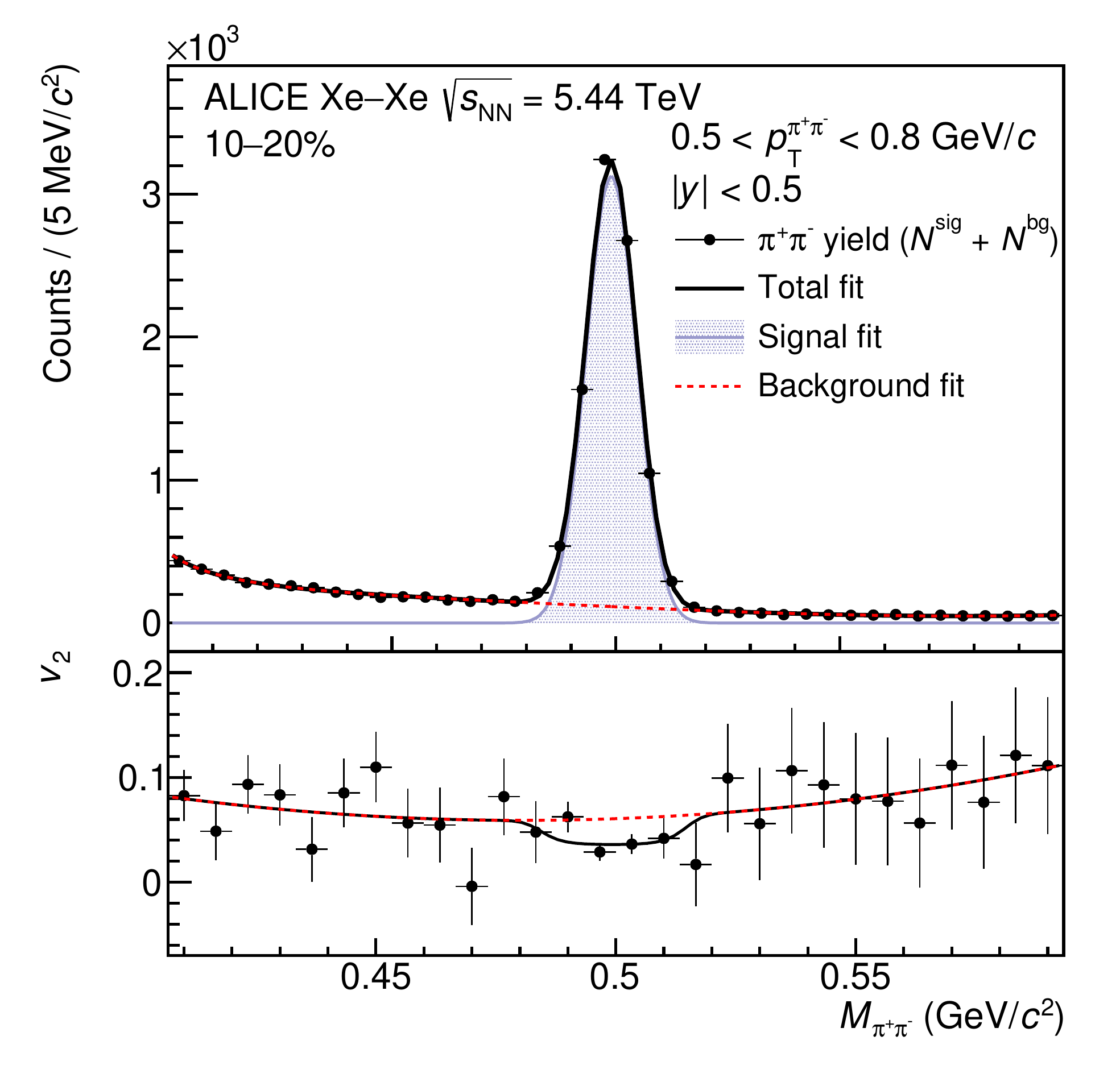}
    \caption{(color online) Top panel: invariant mass distribution of opposite-sign pion pairs belonging to candidate \kzero{} in the centrality range 10--20\% and \pt{} interval $0.5<\pt^{\rm \pi^+\pi^-}<0.8$~\GeVc{}. Bottom panel: a fit of Eq.~\ref{eq:invmassfit} to the mass-dependent $v_2$ distribution.}
    \label{fig:v2K0proc}
    \end{center}
\end{figure}

As the V$^0$s cannot be identified on a track-by-track basis, Eq.~\ref{eq:mth_sp} cannot be used to measure directly $v_{\rm n}$ of \kzero{} and \lambdas{}. Instead, a statistical approach is employed, with the $v_{\rm n}^{\rm{tot}}$ of the candidate V$^0$s being written as the weighted sum of \vnpt{} of the true V$^0$s, $v_{\rm n}^{\rm{sig}}$, and that of the background pairs, $v_{\rm n}^{\rm{bg}}$~\cite{Borghini:2004ra}
\begin{equation}
\label{eq:invmassfit}
    v_{\rm n}^{\rm{tot}} ({M_{\rm{d}^+\rm{d}^-}}) = v_{\rm n}^{\rm{sig}} \frac{N^{\rm{sig}}}{N^{\rm{sig}}+ N^{\rm{bg}}}({M_{\rm{d}^+\rm{d}^-}}) + v_{\rm n}^{\rm{bg}}({M_{\rm{d}^+\rm{d}^-}}) \frac{N^{\rm{bg}}}{N^{\rm{sig}} + N^{\rm{bg}}}({M_{\rm{d}^+\rm{d}^-}}),
\end{equation}
where signal ($N^{\rm sig}$) and background ($N^{\rm bg}$) yields are extracted by integration of the Gaussian distribution and the third-order polynomial function used to parametrize the invariant mass (${M_{\rm{d}^+\rm{d}^-}}$) distribution at the given \pt{}, respectively. The latter accounts for residual contaminations that are present in the \kzero{} and \lambdas{} signals after passing the selection criteria. The $v_{n}^{\rm{tot}}(M_{\rm{d}^+\rm{d}^-})$ obtained according to Eq.~\ref{eq:mth_sp} is fitted using Eq.~\ref{eq:invmassfit} with one parameter for the $v_{\rm n}^{\rm{sig}}$ and a second-order polynomial function to parametrize the $v_{n}^{\rm{bg}}$. This procedure is illustrated in Fig.~\ref{fig:v2K0proc} where the invariant mass distribution of the \kzero{} and a fit of the $v_{2}^{\rm{tot}}({\rm{M}_{\rm{\pi}^+\rm{\pi}^-}})$ distribution are shown in the top and bottom panels, respectively. 

The \pipm{} and p+\pbar{} $v_2$ and $v_3$ are reported for ${0.4<\pt<8.5}$~\GeVc{} and ${0.4<\pt<6.0}$~\GeVc{}, respectively, while \kapm{} $v_{\rm n}$ are presented for ${0.4<\pt<4.0}$~\GeVc{}. The $v_2$ of \kzero{} and \lambdas{} are reported for ${0.5<\pt<6.0}$~\GeVc{} and ${0.8<\pt<6.0}$~\GeVc{}, respectively. All measurements are performed in the rapidity range ${\vert y \vert <}$~0.5.

\section{Systematic uncertainties}
\label{sec:systematics}

\begin{table}[tp]
\centering
\caption{Summary of systematic uncertainties for the $v_2$ of \pipm{}, \kapm{}, p+\pbar{}, \kzero{}, and \lambdas{}. Uncertainties are given as intervals between the minimum and maximum values for all \pt{} and centrality ranges. Empty fields indicate that a given check does not apply, while the field marked \textit{negl.} for negligible implies that the tested uncertainty cannot be resolved within the statistical precision.}
\begin{tabu} to 0.98\columnwidth{ X[5,l] X[1,c] X[1,c] X[1,c] X[1,c] X[1,c] }
Uncertainty source & \pipm{} & \kapm{} & p+\pbar{} & \kzero{} & \lambdas{} \\
\hline											
\hline											
Vertex position & 0--3\% & 0--2\% & 1--3\% & 1--2\% & 1--2\%  \\
1\% wide centrality intervals & 0--2\% & 0--2\% & 0--2\% \\
Centrality estimator & 0--4\% & 0--2\% & 1--4\% & 2--3\% & 1--3\%  \\
Pileup rejection & 0--1\% & 0--1\% & 0--1\% & 0--1\% & 0--1\%  \\
\hline											
Tracking mode & 0--2\% & 0--3\% & 0--5\% \\
Number of TPC space points & 0--1\% & 0--2\% & 0--3\% & 0--1\% & 0--1\% \\
Track quality & 0--1\% & 0--1\% & 0--1\% & 0--2\% & 1--2\%  \\
ITS $\chi^2$ & \textit{negl.} & 0--1\% & 0--1\% \\
\hline											
Particle identification purity & 1--2\% & 1--2\% & 1--3\% & 1--3\% & 1--2\%  \\
Number of TPC clusters used for \dEdx{} & 0--1\% & 0--1\% & 0--1\% & 1--3\% & 1--3\%  \\
Exclusive particle identification & \textit{negl.} & \textit{negl.} & \textit{negl.} \\
\hline											
Decay vertex (radial position) & & & & 1--2\% & 1--4\%  \\
Armenteros--Podolanski variables & & & & 1--2\% 		 \\
DCA decay products to primary vertex & & & & 0--2\% & 1--2\%  \\
DCA between decay products & & & & 1--2\% & 1--2\%  \\
Pointing angle cos $\theta_{\rm p}$ & & & & 0--1\% & \textit{negl.} \\
Minimum \pt{} of daughter tracks & & & & 1--2\% & 0--1\% \\
\dEdx{} contamination for \kzero{}	& 	& 	& 	& 0--2\% 		 \\
\vo{} online selection	& 	& 	& 	& 1--3\% & 0--2\%  \\
Peak shape & & & & 0--1\% & 0--1\%  \\
Residual background in yield & & & & 1--2\% & 0--1\%  \\
\hline											
Positive and negative rapidities & 1--2\% & 1--2\% & 1--3\% & 2--3\% & 1--3\%  \\
Opposite charges & 0--2\% & 0--2\% & 0--2\% \\
$v_{\rm n}^{\rm{bg}}$ parametrization & & & & 0--1\% & 1--2\%  \\
$v_{\rm n}^{\rm{tot}}$ fit ranges & & & & 0--1\% & 0--2\%  \\
\hline											
\end{tabu}
\label{tab:sys1}
\end{table}
The systematic uncertainties are evaluated by varying the event and charged particle tracking selection criteria, the particle identification approach, the \vo{} finding strategy, and the \vnpt{} extraction. The default result is compared to a variation on the nominal measurement. If the value of the variation itself differs from the main result by more than $1\sigma$, which is evaluated based on the recommendations in Ref.~\cite{Barlow:2002yb}, it is considered to be a systematic uncertainty. For various checks performed to quantify the effect of one systematic uncertainty (e.g., using different values for the minimum number of TPC space points employed in the reconstruction to estimate an uncertainty in tracking), the maximum significant deviation found between the nominal measurement and the systematic variations is assigned as a systematic uncertainty. The total systematic uncertainties are estimated by summing in quadrature the systematic uncertainties from the independent sources (if applicable) for all particle species, \vnpt{}, and centrality intervals. A \pt{}-dependent systematic uncertainty is assigned to $v_{\rm n}$ of \pipm, \kapm, and p+\pbar{}, while a \pt{}-independent average uncertainty is reported for $v_2$ of \kzero{} and \lambdas{}. For each particle species, a summary of the magnitude of the relative systematic uncertainties on the values of $v_2$ and $v_3$ are given in Tables~[\ref{tab:sys1}] and~[\ref{tab:sys2}], respectively.

\begin{table}[tp]
\centering
\caption{Summary of systematic uncertainties for the $v_3$ of \pipm{}, \kapm{}, and p+\pbar{}. Uncertainties are given as intervals between the minimum and maximum values for all \pt{} and centrality ranges. The field marked \textit{negl.} for negligible implies that the tested uncertainty cannot be resolved within the statistical precision.}
\begin{tabu} to 0.8\columnwidth{ X[5,l] X[1,c] X[1,c] X[1,c]}
Uncertainty source & \pipm{} & \kapm{} & p+\pbar{} \\
\hline
\hline
Vertex position & 1--3\% & 1--2\% & 1--3\% \\
1\% wide centrality intervals & 0--2\% & 0--2\% & 0--2\% \\
Centrality estimator & 2--4\% & 1--3\% & 2--4\% \\
Pileup rejection & 0--1\% & 0--1\% & 0--1\% \\
\hline
Tracking mode & 0--2\% & 0--4\% & 0--4\% \\
Number of TPC space points & 0--1\% & 0--3\% & 0--2\% \\
Track quality & 0--1\% & 0--1\% & 0--1\% \\
ITS $\chi^2$ & 0--1\% & 0--1\% & 0--1\% \\
\hline
Particle identification purity & 1--3\% & 1--2\% & 2--3\% \\
Number of TPC clusters used for \dEdx{} & 0--2\% & 0--1\% & 0--2\% \\
Exclusive particle identification & \textit{negl.} & \textit{negl.} & \textit{negl.} \\
\hline
Positive and negative rapidities & 1--3\% & 1--2\% & 1--3\% \\
Opposite charges & 0--2\% & 0--2\% & 0--2\% \\
\hline
\end{tabu}
\label{tab:sys2}
\end{table}

Systematic uncertainties related to event selection criteria are estimated by using an alternative centrality estimator based either on the number of hits in the first or second layer of the ITS; by requiring the reconstructed primary vertex position alternatively within $\pm$12~cm, $\pm$7~cm, and $\pm$5~cm from the nominal interaction point along the beam direction; by imposing a stricter pileup rejection than the default selection (i.e., stronger constraints on the consistency of different event multiplicity estimators) or accepting all events with tracks regardless the pileup selection. The limited size of the Xe--Xe data sample does not allow for testing the effects from centrality fluctuations by measuring the $v_{\rm n}$ of \pipm, \kapm, and p+\pbar{} in 1\% wide centrality intervals as done in Refs.~\cite{Acharya:2018ihu, Acharya:2018zuq}. However, the systematic uncertainties estimated for this check in the $v_{\rm n}$ analysis of unidentified charged particles~\cite{Acharya:2018ihu} are applied to the ones for $v_{\rm n}$ of \pipm, \kapm, and p+\pbar{}.

The variations for the track selection criteria are: changing the ITS hit requirements (referred to as tracking mode in Tabs.~\ref{tab:sys1} and~\ref{tab:sys2}); varying the minimum number of TPC space points from 70 to 60, 80, and 90; changing the $\chi^2$ per ITS hit; increasing the minimum number of crossed TPC readout rows from 70 to 120 and the ratio between the number of space points and the number of crossed rows in the TPC from 0.8 to 0.9 (these two checks are combined and referred to as track quality in Tabs.~\ref{tab:sys1} and~\ref{tab:sys2}).

The uncertainties related to particle identification are evaluated by changing the required minimum number of TPC clusters from 70 to 60, 80, and 90 to estimate the effect on the \dEdx{}; varying the maximum value of the ${\rm n}_{\sigma_{\rm PID}}$ from 3 to 1, 2, and 4 for \pt{} $<$ 4~\GeVc{}; rejecting tracks that satisfy the particle identification criterion for more than one particle species simultaneously for \pt{} $<$ 4~\GeVc{}; changing the ${\rm n}_{\sigma_{\rm TPC}}$ ranges for \pt{} $>$ 4~\GeVc{}.

The systematic uncertainty related to the \vo{} finding strategy includes contributions from the topological selection criteria on the \vo{}s themselves and requirements imposed on their daughter tracks. The latter consists of the following variations: requiring in addition $\pt>0.2$ \GeVc{} for each daughter track; changing the minimum number of TPC space points from 70 to 60 and 80; varying the minimum number of crossed TPC readout rows from 70 to 60 and 80; increasing the ratio between the number of space points and the number of crossed rows in the TPC from 0.8 to 0.9; varying the minimum DCA of the \vo{} daughter tracks to the primary vertex from 0.1~cm to 0.05~cm and 0.3~cm; changing the maximum DCA of the \vo{} daughter tracks to the secondary vertex from 0.5~cm to 0.3~cm and 0.7~cm; requesting at least 60 and 90 TPC clusters instead of 70 to estimate the effect on the \dEdx{}; varying the maximum absolute value of the ${\rm n}_{\sigma_{\rm TPC}}$ from 3 to 1 and 4. Concerning the V$^0$s selection, the following variations are investigated: changing the minimum value of the $\cos \theta_{\rm p}$ from 0.998 to 0.98; requesting a minimum radial distance to the beam line at which the \vo{} can be produced of 1~cm and 15~cm instead of 5~cm; changing the maximum radial distance to the beam pipe at which the \vo{} can be produced from 100~cm to 50~cm and 150~cm; suppressing the contamination from \lambdas{} and electron--positron pairs coming from $\gamma$ conversions to the \kzero{} sample by limiting the value of the Armenteros--Podolanski variables and excluding electrons by only selecting \vo{} daughter tracks with a \dEdx{} value 2$\sigma$ away from the expected electron \dEdx{}. Finally, the yield extraction is varied by using polynomials of different orders as parametrization of the residual background in the invariant mass spectra and employing a sum of two Gaussian distributions with the same mean for the parametrization of the \kzero{} and \lambdas{} invariant mass yield. 

The uncertainties associated with the determination of \vnpt{} are estimated by performing the analysis for positive and negative rapidities independently, thus increasing the pseudorapidity gap to $|\Delta\eta|>2.8$; performing the analysis for \pipm{}, \kapm{}, and p+\pbar{} for positive and negative charges independently; varying the $M_{\rm d^+ d^-}$ range over which Eq.~\ref{eq:invmassfit} is fitted; changing the $v_2^{\rm bg}$ parametrization from a second-order polynomial to a linear or constant function.

\section{Results and discussion}
\label{sec:results}

\subsection{Centrality and \texorpdfstring{\pt\,}~dependence of flow coefficients}
\begin{figure}[tp]
  \centering
  \includegraphics[width=\textwidth]{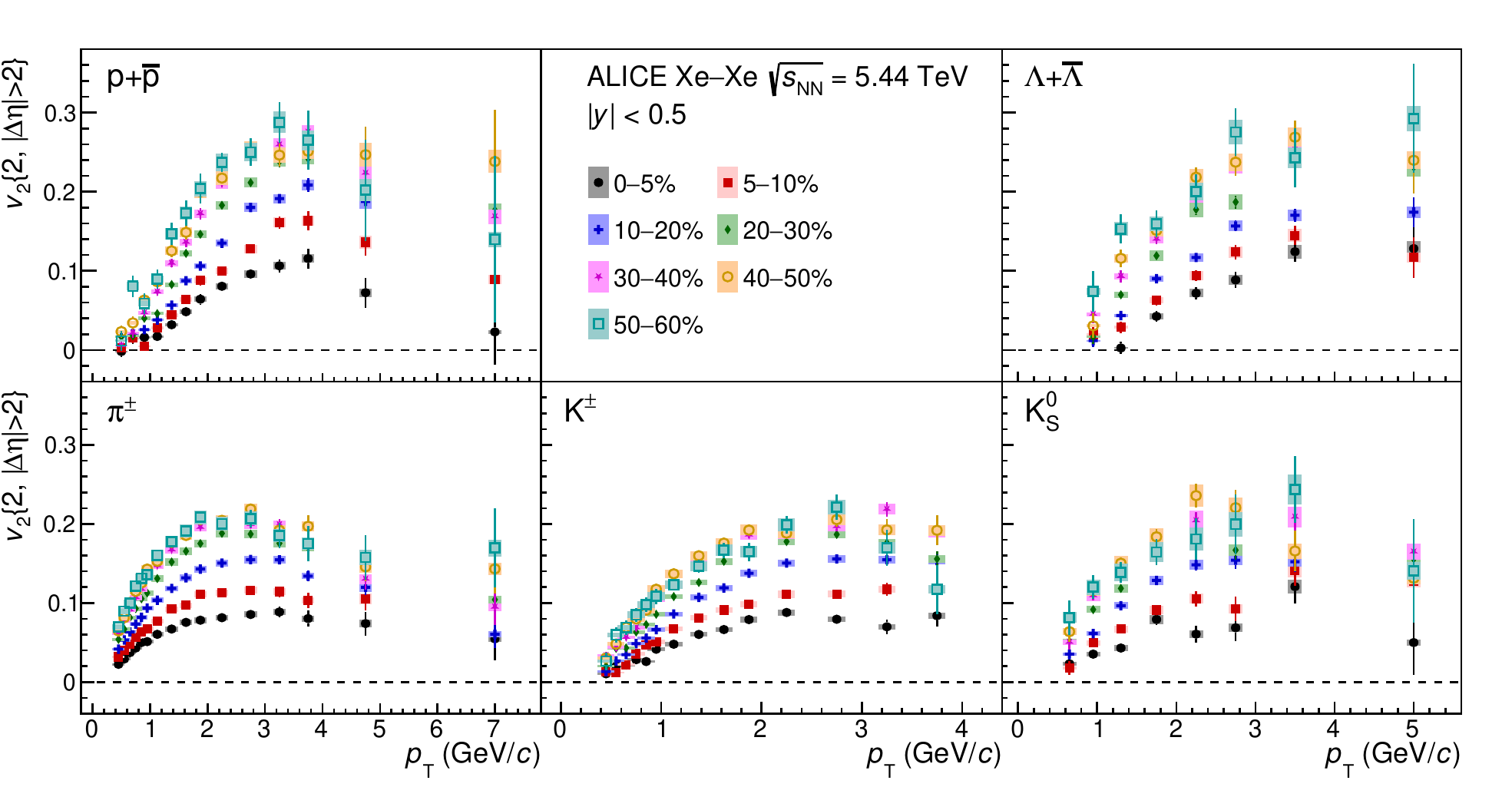}
  \caption{(color online) Centrality dependence of \vtwopt{} for \pipm{}, \kapm{}, p+\pbar{}, \kzero{}, and \lambdas{}. Bars (boxes) denote statistical (systematic) uncertainties.}
  \label{fig:v2pidcen}
\end{figure}
\begin{figure}[tp]
  \centering
  \includegraphics[width=\textwidth]{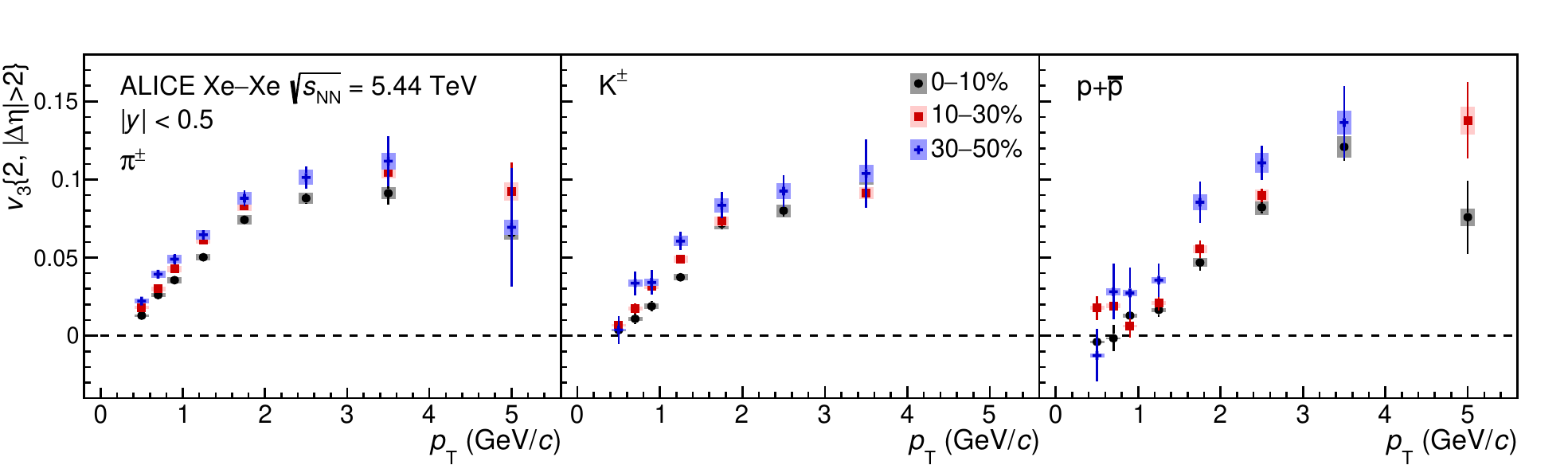}
  \caption{(color online) Centrality dependence of \vthreept{} for \pipm{}, \kapm{}, and p+\pbar{}. Bars (boxes) denote statistical (systematic) uncertainties.}
  \label{fig:v3pidcen}
\end{figure}

The \vtwopt{} of \pipm{}, \kapm{}, p+\pbar{}, \kzero{}, and \lambdas{} is presented in Fig.~\ref{fig:v2pidcen} for various centrality intervals in the \mbox{0--60\%} range. The measured $v_2$ of all particle species, being mainly driven by the collision geometry, increases strongly with decreasing centrality up to the 40--50\% centrality interval. This evolution is expected since $v_2$ scales approximately linearly with the eccentricity of the overlap zone of the colliding nuclei~\cite{Gardim:2011xv}. For the 50--60\% centrality class, the value of $v_2$ is similar to that measured in the previous centrality interval within uncertainties, which is expected due to a shorter lifetime of the system in more peripheral collisions. This together with the reduced contribution of eccentricity fluctuations and hadronic interactions inhibit the generation of large $v_2$~\cite{Song:2007fn, Song:2013qma}. The \vtwopt{} increases up to $\pt \sim$ 3--4 \GeVc{}, where a maximum is reached, and then decreases with increasing \pt{}. The position of this maximum depends weakly on centrality and is located at smaller \pt{} for lighter compared to heavier particles, over the various centrality intervals studied. The observed phenomenon finds an explanation in the changes in parton density and the centrality dependence of radial flow~\cite{Acharya:2018zuq}, which will be detailed in Sec.~\ref{sec:shapeV2}. The evolution of $v_2$ with \pt{} and centrality is similar to that reported in Pb--Pb collisions~\cite{Abelev:2014pua, Adam:2016nfo, Acharya:2018zuq}.

Unlike $v_2$, the third-order flow coefficient $v_3$ originates from event-by-event fluctuations in the initial nucleon density distribution \cite{Bhalerao:2006tp, Alver:2008zza, Takahashi:2009na, Alver:2010gr, Alver:2010dn}. A stronger decrease of $v_3$ compared to $v_2$ is expected due to the dampening effect of $\eta/s$, which implies that $v_3$ is more sensitive to transport coefficients than $v_2$ \cite{Qin:2010pf, Teaney:2010vd}. The limited size of the Xe--Xe data sample does not allow for $v_3$ to be measured accurately in the centrality intervals used for $v_2$. Therefore, these measurements have been combined in larger centrality classes using the \pt-differential yields~\cite{Acharya:2021ljw} as weights. Figure~\ref{fig:v3pidcen} presents the \vthreept{} of \pipm{}, \kapm{}, and p+\pbar{} for the 0--10\%, 10--30\%, and 30--50\% centrality intervals. The measured $v_3$ is non-zero, positive for most of the \pt{} ranges and increases with \pt{} up to 3--4 \GeVc{}. The coefficient $v_3$ shows a weak centrality dependence with a magnitude significantly smaller than that of $v_2$, except for the 0--10\% centrality interval. These findings illustrate that $v_3$ originates from fluctuations of the initial geometry of the system.
\begin{figure}[tp]
  \centering
  \includegraphics[width=0.85\textwidth]{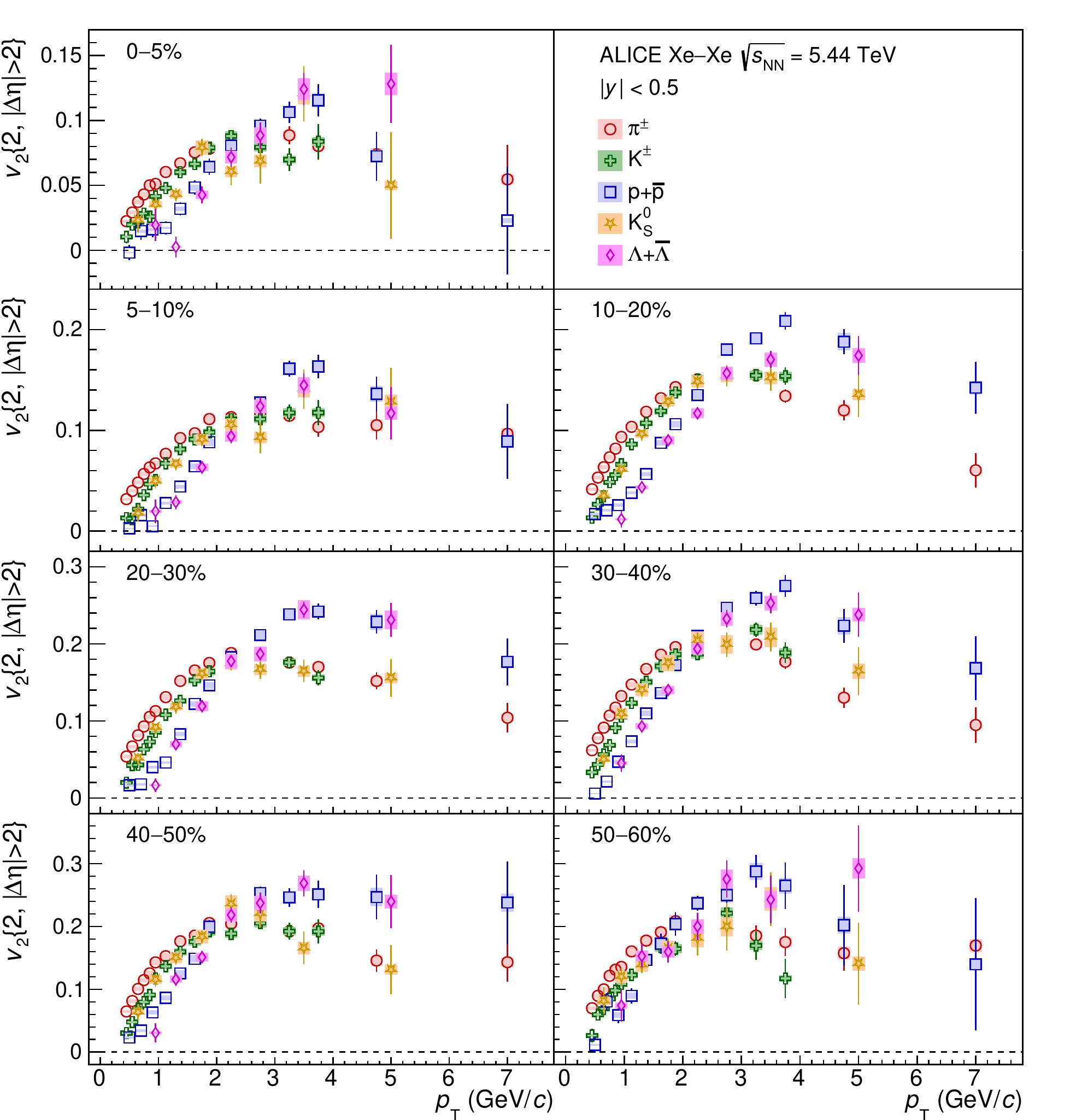}
  \caption{(color online) The \pt-differential $v_2$ of \pipm{}, \kapm{}, p+\pbar{}, \kzero{}, and \lambdas{} in a given centrality interval. Bars (boxes) denote statistical (systematic) uncertainties.}
  \label{fig:v2pid}
\end{figure}
\begin{figure}[!t]
  \centering
  \includegraphics[width=\textwidth]{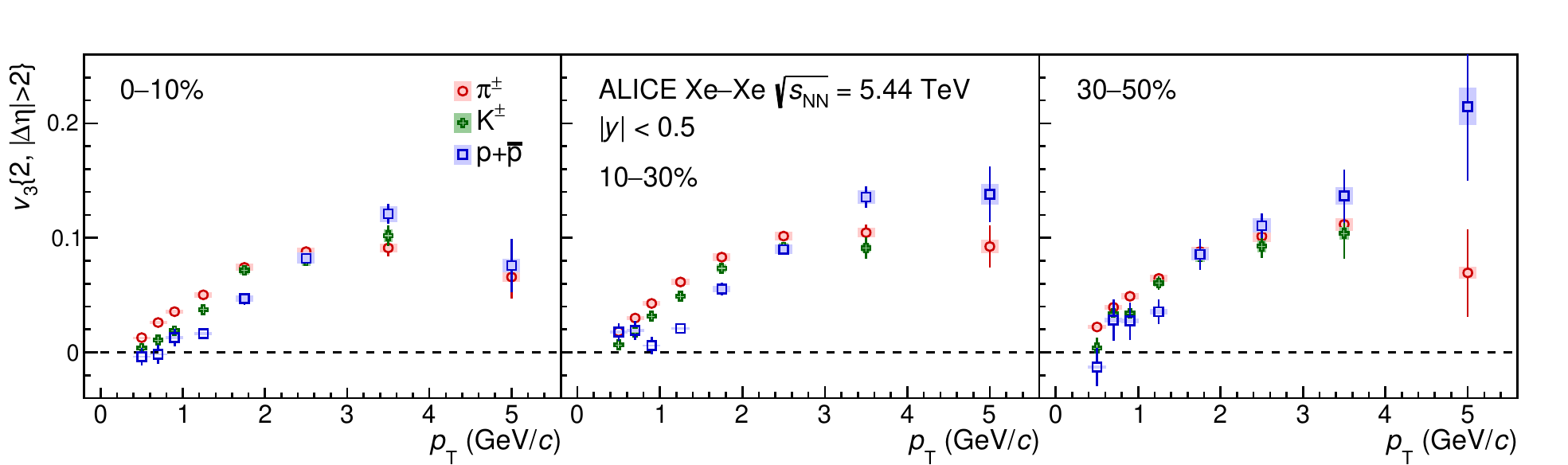}
  \caption{(color online) The \pt-differential $v_3$ of \pipm{}, \kapm{}, and p+\pbar{} in a given centrality interval. Bars (boxes) denote statistical (systematic) uncertainties.}
  \label{fig:v3pid}
\end{figure}

Figure~\ref{fig:v2pid} shows comparisons of the \vtwopt{} for all particle species in a given centrality interval arranged into panels of various centrality classes. For $\pt<$ 2--3~\GeVc{}, $v_2$ of the different particle species exhibits a mass ordering, meaning that heavier particles have a smaller $v_2$ than that of lighter particles at the same \pt{}. This behaviour can be attributed to the interplay of elliptic flow with radial flow which imposes an isotropic velocity boost equal for all particles, thus pushing heavier particles towards higher $\pt$~\cite{Huovinen:2001cy, Shen:2011eg}. For $3<\pt<$ 8 \GeVc{}, the $v_2$ of baryons becomes larger than that of mesons, indicating that the particle type dependence persists out to high \pt{}. This grouping according to the number of constituent quarks supports the hypothesis of particle production via quark coalescence~\cite{Molnar:2003ff}. The crossing between meson and baryon $v_2$ depends on particle species and centrality, occurring at lower \pt{} values for peripheral than central collisions as a result of the smaller radial flow in the former. Comparing the \kapm{} and \kzero{} $v_2$, there is a hint of $v_2^{\rm \kzero{}} < v_2^{\rm \kapm{}}$ in the 0--10\% centrality range, while the measurements are compatible within statistical uncertainties in the 10--60\% centrality interval. One should note that a difference in \vtwopt{} of \kapm{} and \kzero{} was reported by ALICE in Pb--Pb collisions~\cite{Abelev:2014pua, Acharya:2018zuq}.

Figure~\ref{fig:v3pid} presents the \vthreept{} of \pipm{}, \kapm{}, and p+\pbar{} in a given centrality interval. The $v_3$ of different particle species is mass ordered at $\pt<$ 2--3~\GeVc{}, indicating the interplay between triangular and radial flow. For $3<\pt<6$~\GeVc{}, the p+\pbar{} $v_3$ is slightly larger than that of \pipm{}. The crossing between $v_3$ values of pions and protons shows a weak centrality dependence.

\subsection{Scaling properties}
\begin{figure}[tp]
  \centering
  \includegraphics[width=0.85\textwidth]{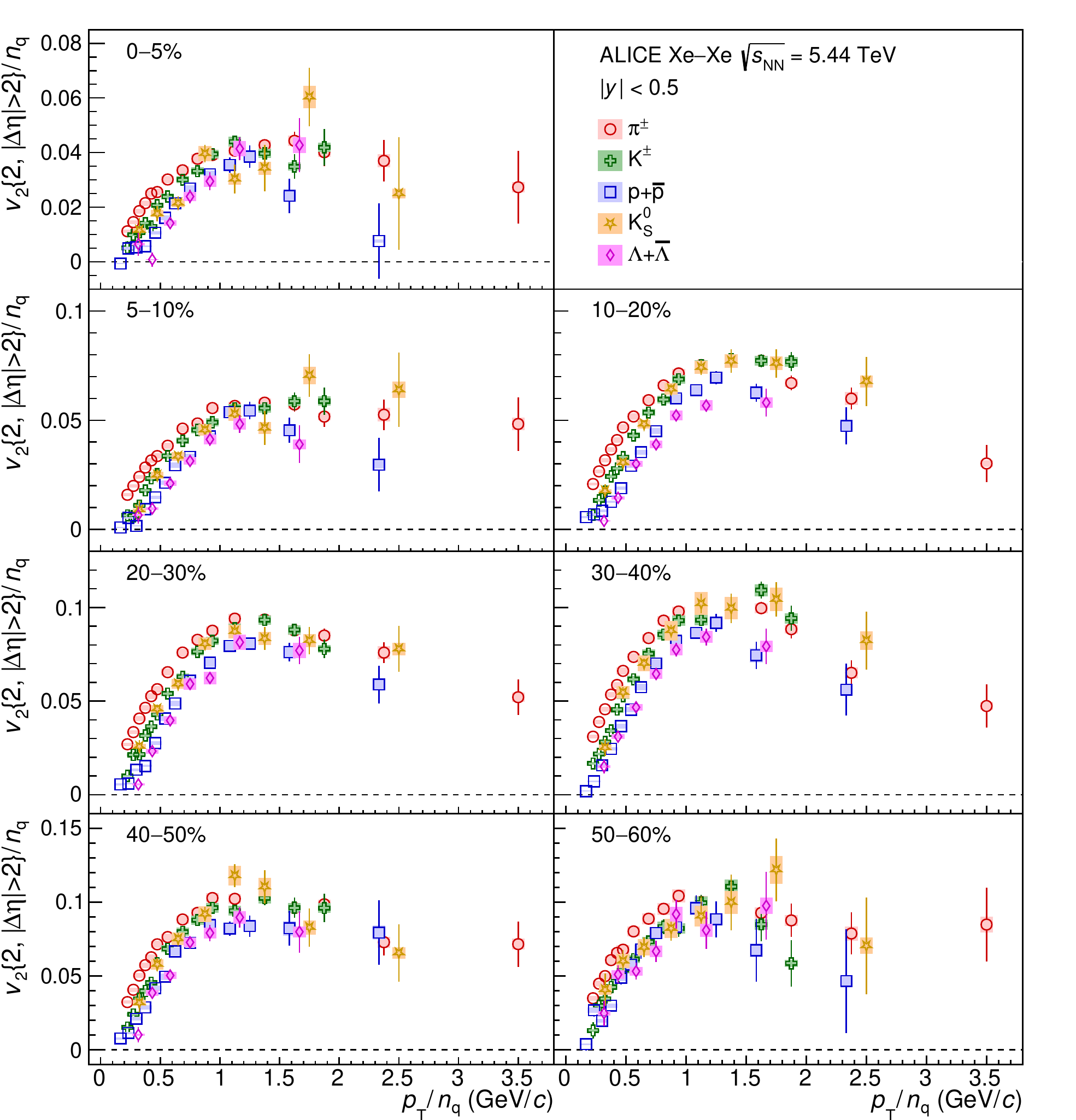}
  \caption{(color online) The $\pt/n_{\rm q}$ dependence of $v_2/n_{\rm q}$ of \pipm{}, \kapm{}, p+\pbar{}, \kzero{}, and \lambdas{} for various centrality classes. Bars (boxes) denote statistical (systematic) uncertainties.}
  \label{fig:v2nq}
\end{figure}
\begin{figure}[!t]
  \centering
  \includegraphics[width=\textwidth]{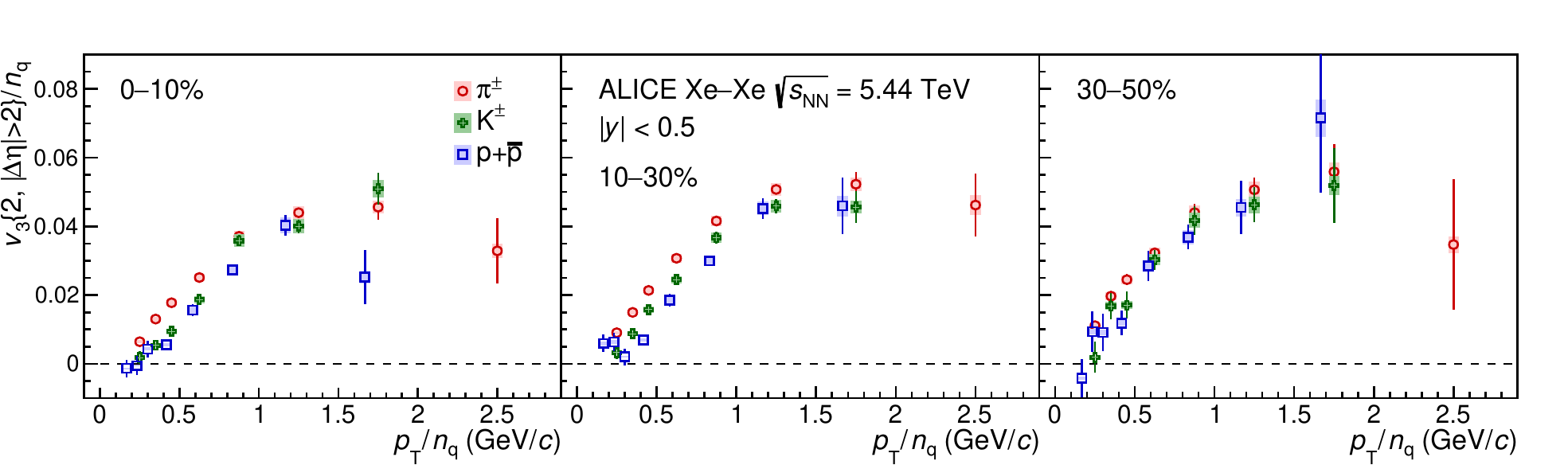}
  \caption{(color online) The $\pt/n_{\rm q}$ dependence of $v_3/n_{\rm q}$ of \pipm{}, \kapm{}, and p+\pbar{} for various centrality classes. Bars (boxes) denote statistical (systematic) uncertainties.}
  \label{fig:v3nq}
\end{figure}

Scaling with the number of constituent quarks (NCQ) of $v_{\rm n}$ has been suggested to test the hypothesis of particle production via quark coalescence at intermediate \pt{}, which would lead to a meson and baryon $v_{\rm n}$ grouping~\cite{Molnar:2003ff, Greco:2003mm, Fries:2003kq}. This can be achieved by dividing both $v_{\rm n}$ and \pt{} by the number of constituent quarks ($n_{\rm q}$) independently for each particle species. Figures~\ref{fig:v2nq} and~\ref{fig:v3nq} present the $v_{2}/n_{\rm q}$ and $v_{3}/n_{\rm q}$ as function of $\pt/n_{\rm q}$ for \pipm{}, \kapm{}, p+\pbar{}, \kzero{}, and \lambdas{}, for various centrality classes. For $1 < \pt/n_{\rm q} < 3$~\GeVc{}, the region where quark coalescence is hypothesized to be the dominant process~\cite{Molnar:2003ff, Greco:2003mm}, a deviation from the exact scaling of $\pm$ 20\% is found for $v_2$, similar to the one reported in Pb--Pb collisions~\cite{Abelev:2014pua, Adam:2016nfo, Acharya:2018zuq}. This deviation is quantified by dividing the $\pt/n_{\rm q}$ dependence of $v_2/n_{\rm q}$ by a cubic spline fit to the p+\pbar{} $v_2/n_{\rm q}$. The scaling for $v_3$ seems to hold within the relatively large uncertainties.

\subsection{Shape evolution of \texorpdfstring{\vtwopt \,}~as function of centrality}
\label{sec:shapeV2}
\begin{figure}[tp]
    \begin{center}
    \includegraphics[width=0.85\textwidth]{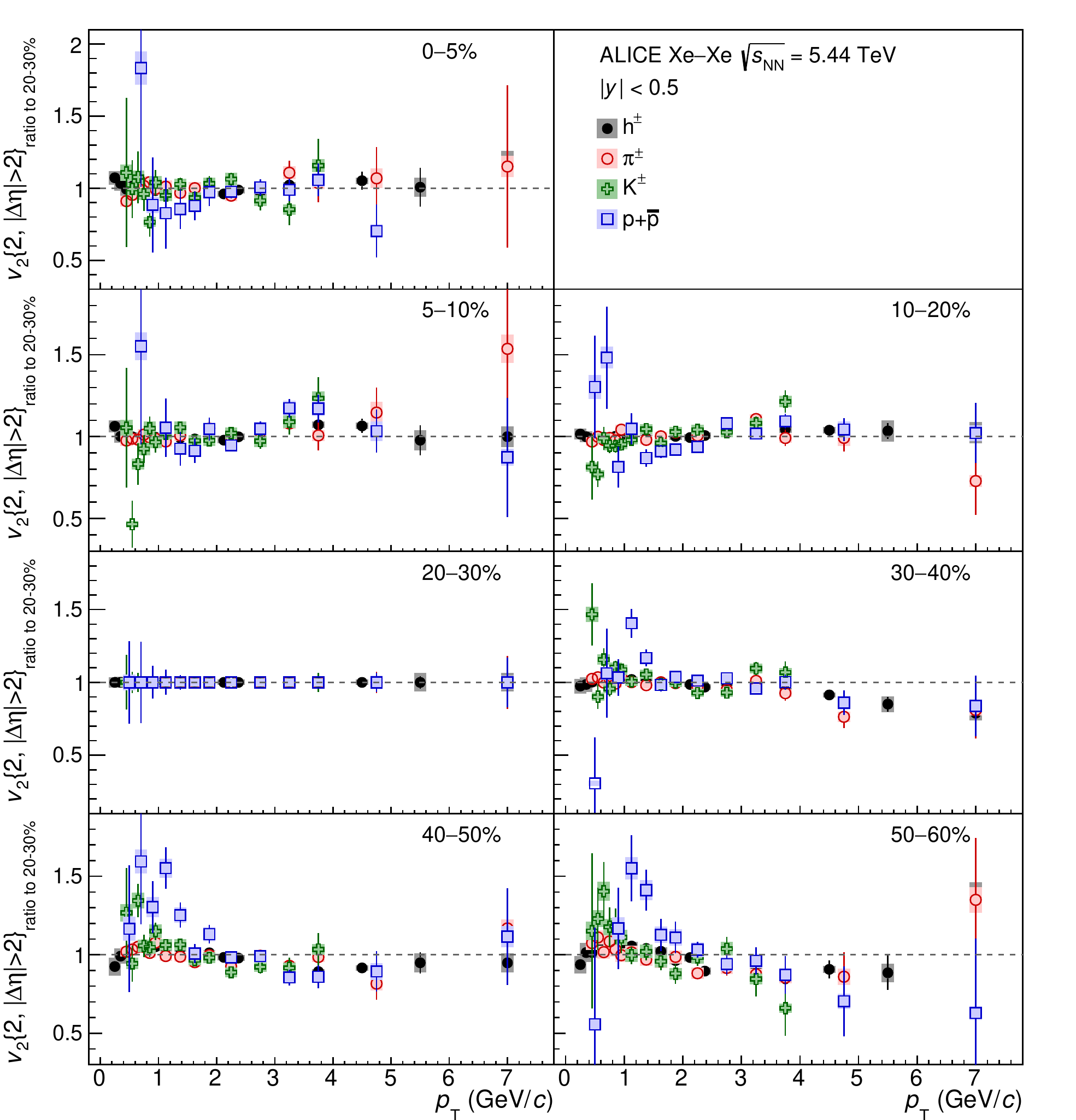}
    \caption{(color online) Centrality dependence of $\vtwopt{}_{\rm ratio~to~20-30\%}$ for \pipm{}, \kapm{}, p+\pbar{}, and inclusive charged hadrons (h$^{\pm}$)~\cite{Acharya:2018ihu}. Bars (boxes) denote statistical (systematic) uncertainties.}
    \label{fig:v2Shape}
    \end{center}
\end{figure}

The centrality dependence of the shape evolution of \vtwopt{} is studied as in Ref.~\cite{Acharya:2018zuq} by choosing the $v_2$ measured in the 20--30\% centrality interval as reference. It is quantified by dividing the \vtwopt{} in a given centrality interval by this reference and denoted as $\vtwopt{}_{\rm ratio~to~20-30\%}$ in the following. The ratio of the \pt{}-integrated $v_2$ value obtained in the 20--30\% centrality interval to that in the centrality interval of interest is used as a normalization factor in order for $\vtwopt{}_{\rm ratio~to~20-30\%}$ to be unity in the absence of centrality-dependent variations. The shape evolution of elliptic flow for \pipm{}, \kapm{}, p+\pbar{}, and inclusive charged hadrons (the latter taken from Ref.~\cite{Acharya:2018ihu}) is presented in Fig.~\ref{fig:v2Shape}. Variations in shape of about 10\% are observed for inclusive charged hadrons throughout the considered \pt{} range within uncertainties. The evolution of the shape of the \vtwopt{} shows different trends for \pipm{}, \kapm{}, and p+\pbar{} for $\pt<2$ \GeVc{} and no particle type dependence within uncertainties for $\pt \ge 2$ \GeVc{}. The variations are more pronounced for p+\pbar{} $\vtwopt{}_{\rm ratio~to~20-30\%}$, reaching around 60\% at low \pt{} in peripheral collisions. The elliptic flow of \kapm{} varies up to 40\% for $\pt<1$ \GeVc{}, while the $v_2(\pt)_{\rm ratio~to~20-30\%}$ of \pipm{} follows the results for inclusive charged particles. Radial flow and transverse quark density should play important roles in this mass dependence for $\pt<2$ \GeVc{} as both depend on centrality, having larger values in central than peripheral collisions. The latter influences the peak value of \vnpt{} in the coalescence model~\cite{Fries:2008hs}, while the effect of the former on $v_{\rm n}$ of heavier particles is greater than on the lighter particles at low \pt{}.
\begin{figure}[tp]
    \begin{center}
        \includegraphics[width=.69\textwidth]{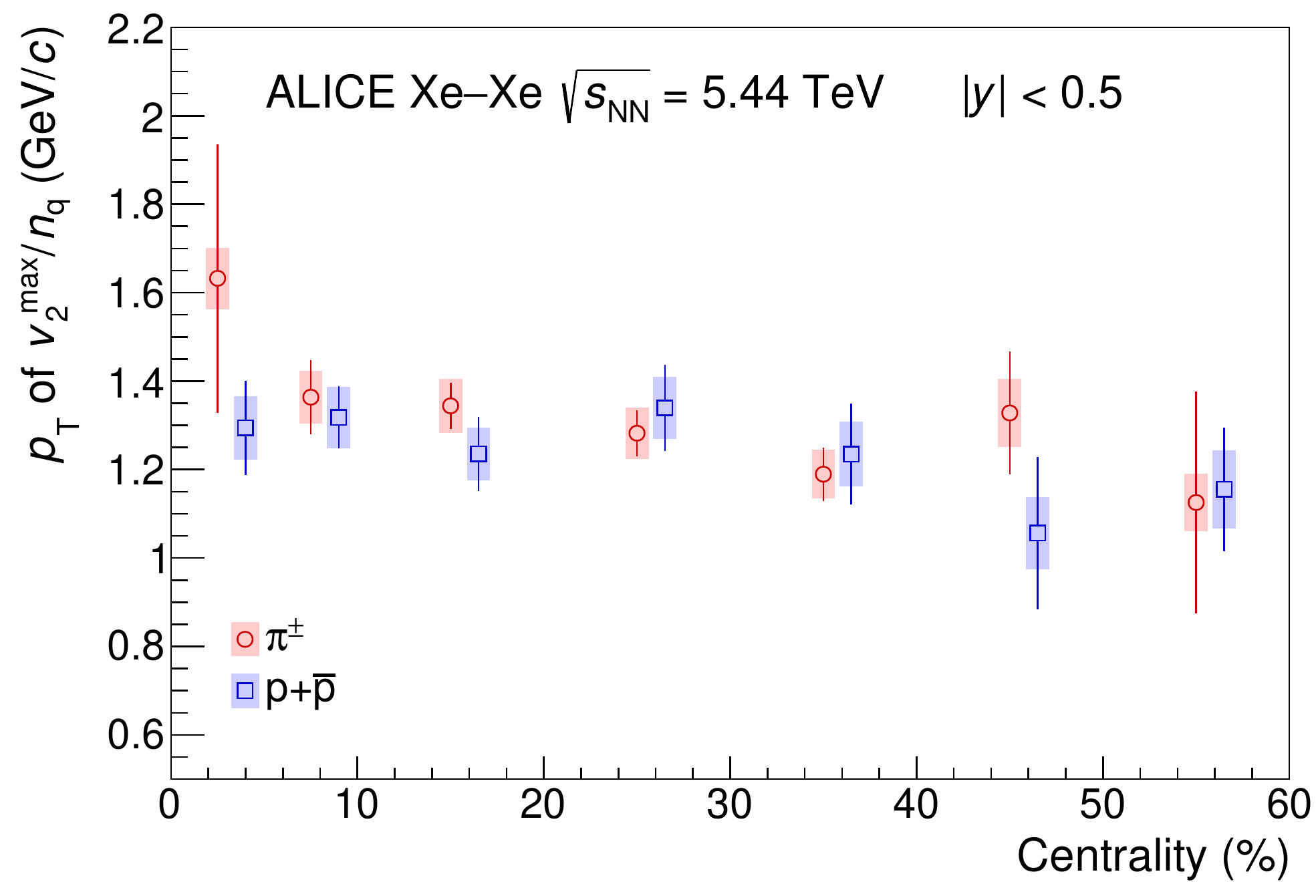}
        \caption{(color online) Centrality dependence of \vtwomax{} for \pipm{} and p+\pbar{} divided by number of constituent quarks, $n_{\rm q}$. The p+\pbar{} points are slightly shifted along the horizontal axis for better visibility. Bars (boxes) denote statistical (systematic) uncertainties.}
    \label{fig:v2maxpt}
    \end{center}
\end{figure}
\begin{figure}[tbp]
    \begin{center}
    \includegraphics[width=\textwidth]{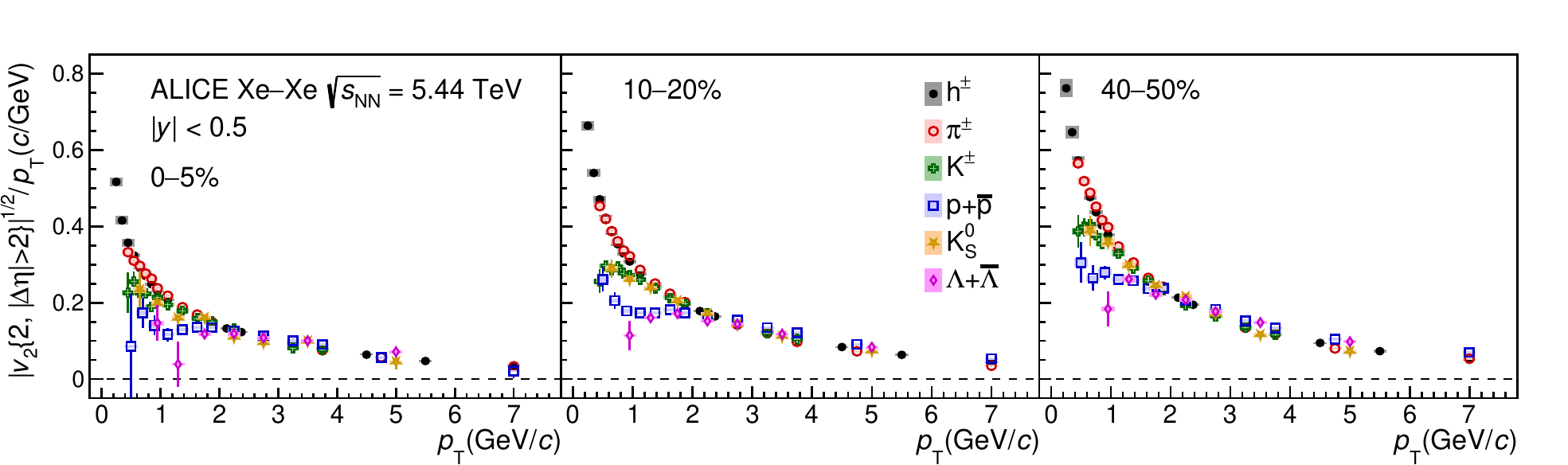}
    \caption{(color online) $|v_2|^{1/2} / \pt{}$ of inclusive charged hadrons (h$^{\pm}$) \cite{Acharya:2018ihu}, \pipm{}, \kapm{}, p+\pbar{}, \kzero{}, and \lambdas{} as function of \pt{} for various centrality intervals. Bars (boxes) denote statistical (systematic) uncertainties.}
    \label{fig:v2ratpt}
    \end{center}
\end{figure}

An alternative way of quantifying the shape of the \vtwopt{} is the position of the maximum $v_2$. It is expected to be located at higher \pt{} in central than peripheral collisions as the quark density depends on centrality. Its centrality dependence, quantified by the \pt{} where \vtwopt{} reaches a maximum divided by the number of constituent quarks $n_{\rm q}$, is reported in Fig.~\ref{fig:v2maxpt} for \pipm{} and p+\pbar{}. The \kapm{}, \kzero{}, and \lambdas{} are not included since the kinematic range and granularity of the measurements do not allow for a reliable extraction of a maximum. The \pt{}$/n_{\rm q}$ at which \vtwopt{} reaches a maximum, denoted as \vtwomax{}, shows a weak centrality dependence with a decreasing trend from central to peripheral collisions. This behavior is expected from the hypothesis of hadronization through coalescence where an increase in the transverse density of quarks, as in more central collisions, results in a higher value of \vtwomax{}~\cite{Fries:2008hs}. The observed \vtwomax{} is compatible between \pipm{} and p+\pbar{} for all centrality intervals within uncertainties. The systematic uncertainties presented in Fig.~\ref{fig:v2maxpt} are evaluated directly on \vtwomax{} to accurately take into account that some systematic uncertainties can be point-by-point correlated in \pt{}.
 
If $v_2$ exhibits a power law dependence on $\pt^2$ up to $\pt{} \sim M$ for particles with mass $M$ as in the scenario of ideal hydrodynamics~\cite{Borghini:2005kd}, ratios of the form $\vert v_2\vert^{1/2}/ \pt{}$ should be constant. Previous measurements performed by ALICE in Pb--Pb collisions~\cite{Acharya:2018zuq} have shown that the $v_2 \propto \pt{}^2$ scaling is broken for \pipm{} and the inclusive charged particles for all centrality intervals. However, this scaling holds up to $\pt{}~\approx~1$~\GeVc{} for \kapm{} and \kzero{}, and up to $\pt{}~\approx~2$~\GeVc{} for p+\pbar{} and \lambdas{} for central and semicentral collisions~\cite{Acharya:2018zuq}. It should be noted, however, that the kinematic constraints imposed on the measurement preclude testing the scaling hypothesis in the full relevant momentum region for \pipm{} and the inclusive charged particles. Figure~\ref{fig:v2ratpt} shows $\vert v_2\vert^{1/2}/ \pt{}$ for inclusive charged particles~\cite{Acharya:2018ihu}, \pipm{}, \kapm{}, p+\pbar{}, \kzero{}, and \lambdas{} as a function of \pt{} in various centrality intervals. The ratios $\vert v_2\vert^{1/2}/ \pt{}$ show a strong \pt{} dependence for \pipm{} and the inclusive charged particles, while they exhibit a weak (if any) \pt{} dependence up to $\pt{}~\approx~1$~\GeVc{} for \kapm{} and \kzero{}, and up to $\pt{}~\approx~2$~\GeVc{} for p+\pbar{} and \lambdas{} for the 0--5\% and 10--20\% centrality intervals.

\subsection{Comparison with hydrodynamic calculations}
\label{sec:hydroComp}
\begin{figure}[tp]
    \begin{center}
    \includegraphics[width=0.99\textwidth]{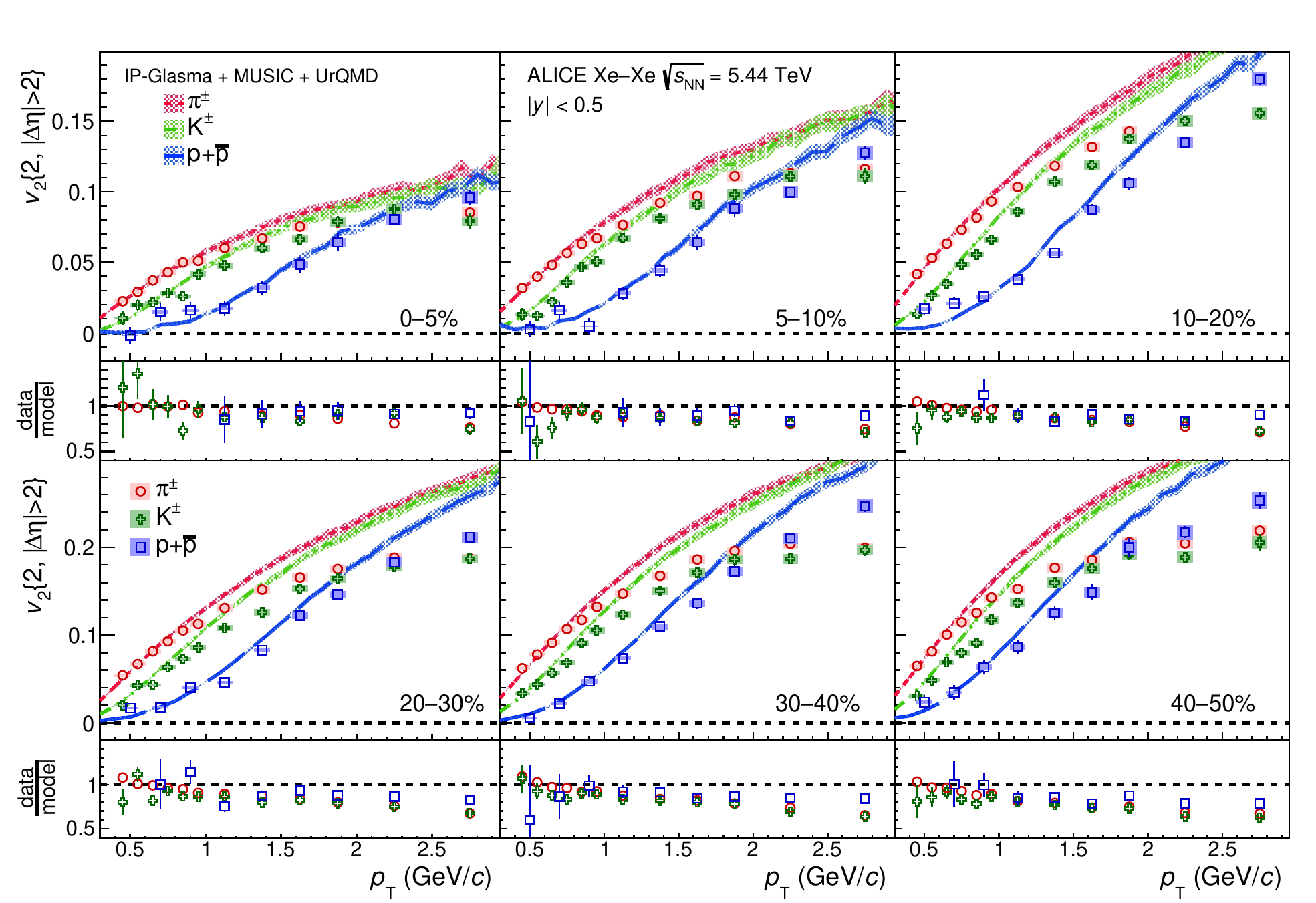}
    \caption{(color online) The $\pt$-differential $v_2$ of \pipm{}, \kapm{}, and p+\pbar{} for various centrality classes compared to hydrodynamic calculations from MUSIC model using IP-Glasma initial conditions (colored curves)~\cite{Schenke:2020mbo}. Bars (boxes) denote statistical (systematic) uncertainties. The uncertainties of the hydrodynamic calculations are depicted by the thickness of the curves. The ratios of the measured $v_2$ to a fit to the hydrodynamic calculations are also presented for clarity.}
    \label{fig:v2_compIPGM}
    \end{center}
\end{figure}

Figure~\ref{fig:v2_compIPGM} presents the $\pt$-differential $v_2$ of \pipm{}, \kapm{}, and p+\pbar{} for various centrality intervals compared with predictions from MUSIC hydrodynamic simulations~\cite{Schenke:2020mbo}. MUSIC~\cite{Schenke:2010rr}, an event-by-event 3+1 dimensional viscous hydrodynamic model, uses the IP-Glasma model~\cite{glasma1, glasma2} to describe the initial conditions of the collision and is coupled to a hadronic cascade model (UrQMD)~\cite{glasmacascade,bleichercascade}, which allows one to study the influence of the hadronic phase on the development of anisotropic flow for different particle species. The starting time for the hydrodynamic evolution and the switching energy between hydrodynamics and the microscopic transport evolution are set to $\tau_0=0.4$~fm/$c$ and $e_{\rm sw}= 0.18$~GeV/fm$^3$, respectively. A value of $\eta/s = 0.12$ and a temperature dependent $\zeta/s$ are also employed in this model. It should be noted that these parameters do not depend on collision system or centrality.% a switching temperature $T_{\rm sw}=145$~MeV

Figure~\ref{fig:v2_compIPGM} shows that the MUSIC calculations qualitatively reproduce the mass ordering. The predictions are in agreement with the measured \vtwopt{} of \pipm{}, \kapm{}, and p+\pbar{} for $\pt{}<1$~\GeVc{}, while they overestimate the data points at higher \pt{}. However, the $v_2$ of p+\pbar{} is more accurately described than that of \pipm{} and \kapm{} for ${\pt{} \ge 1}$~\GeVc{} in all centrality intervals. A better agreement with the data points is found in central than in peripheral collisions. The differences between the data points and model are also illustrated in Fig.~\ref{fig:v2_compIPGM} as the ratios of the measured $v_2$ to a fit to the theoretical calculations.

\subsection{Comparison with \texorpdfstring{$v_{\rm n}$ \,}~of identified particles in Pb--Pb collisions at \texorpdfstring{$\snn = 5.02$ \,}~TeV}
\begin{sidewaysfigure} 
  \centering
  \includegraphics[width=0.98\textwidth,  height=13cm]{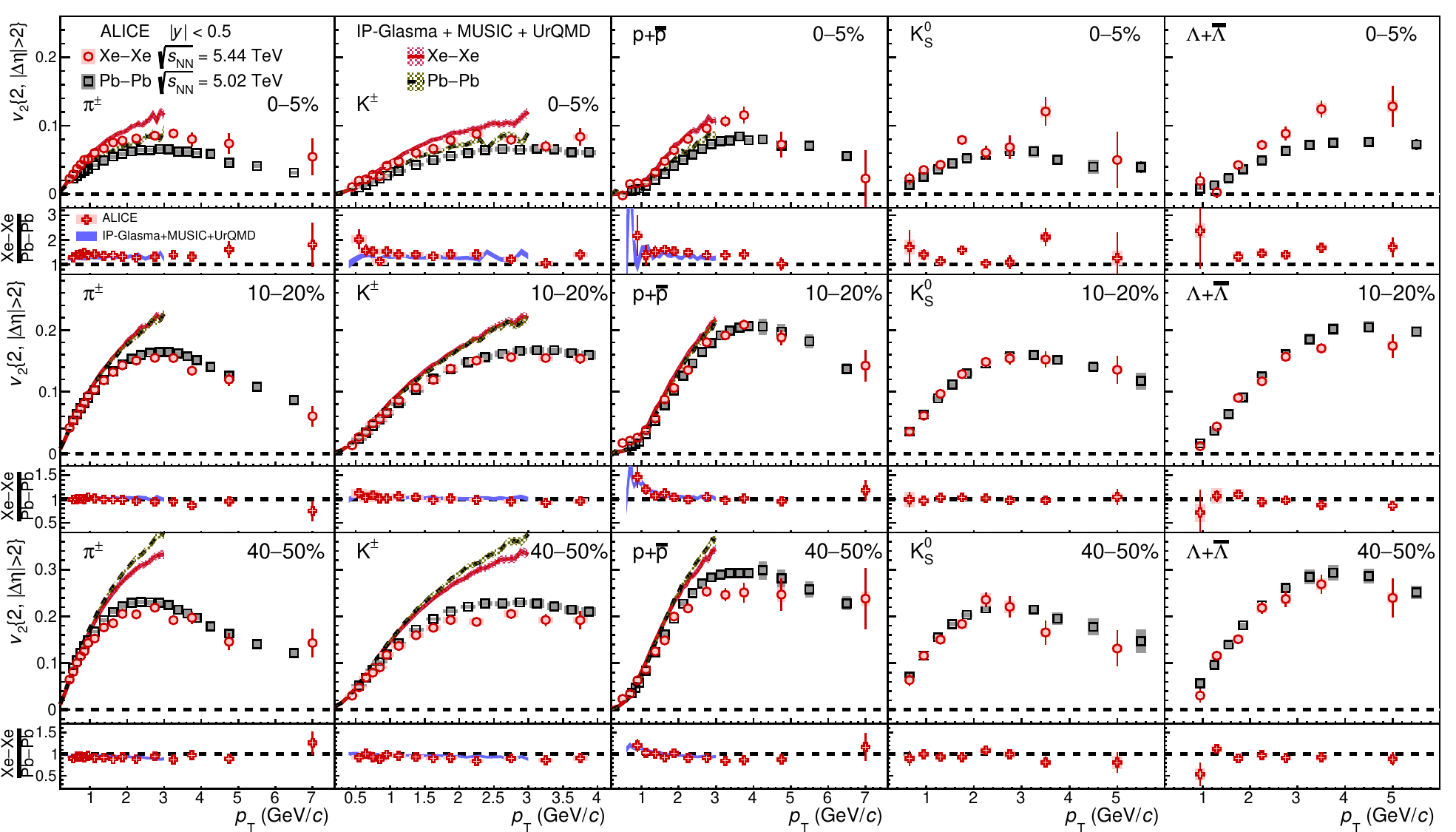}
  \caption{(color online) The $\pt$-differential $v_2$ of \pipm{}, \kapm{}, p+\pbar{}, \kzero{}, and \lambdas{} (red markers) compared to ALICE measurements performed in Pb--Pb collisions at $\sqrt{s_{\rm NN}}$~=~5.02 TeV~\cite{Acharya:2018zuq} (black markers) for the 0--5\% (top panels), 10--20\% (middle panels), and 40--50\% (bottom panels) centrality intervals. The ratios of Xe--Xe measurements to a cubic spline fit to Pb--Pb measurements are also presented for clarity. The colored curves represent hydrodynamic calculations from MUSIC model using IP-Glasma initial conditions~\cite{Schenke:2020mbo}. Bars (boxes) denote statistical (systematic) uncertainties. The uncertainties of the hydrodynamic calculations are depicted by the thickness of the curves.}
  \label{fig:v2_Xe_vs_Pb}
\end{sidewaysfigure}
\begin{figure}[tp]
  \centering
  \includegraphics[width=\textwidth]{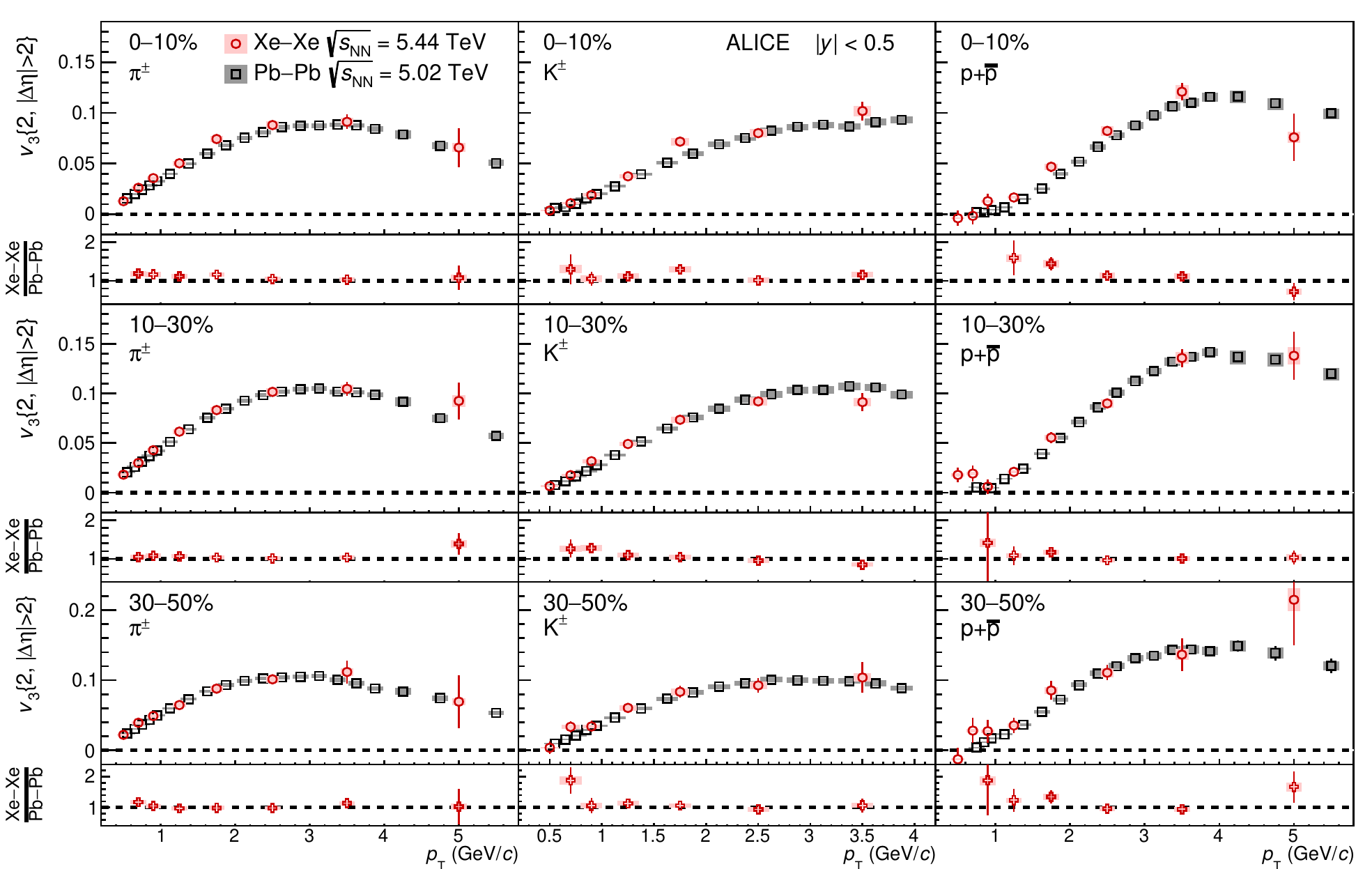}
  \caption{(color online) The $\pt$-differential $v_3$ of \pipm{}, \kapm{}, and p+\pbar{} (black markers) compared to ALICE measurements performed in Pb--Pb collisions at $\sqrt{s_{\rm NN}}$~=~5.02 TeV~\cite{Acharya:2018zuq} (red markers) for the 0--10\% (top panels), 10--30\% (middle panels), and 30--50\% (bottom panels) centrality classes. The ratios of Xe--Xe measurements to a cubic spline fit to Pb--Pb measurements are also presented for clarity. Bars (boxes) denote statistical (systematic) uncertainties.}
  \label{fig:v3_Xe_vs_Pb}
\end{figure}
\begin{figure}[htbp]
  \centering
  \includegraphics[width=0.98\textwidth]{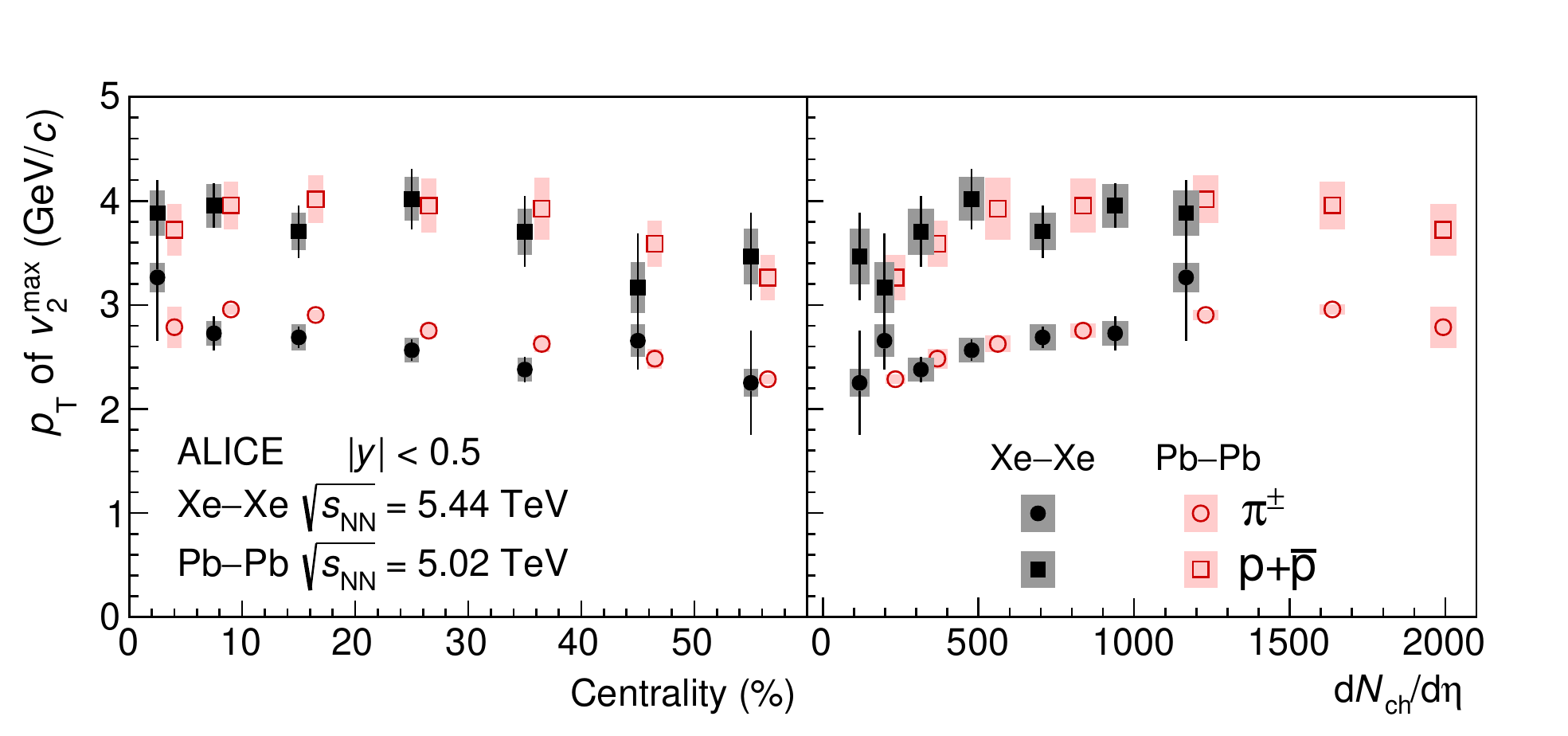}
  \caption{(color online) The \vtwomax{} for \pipm{} and p+\pbar{} (black markers) compared to ALICE measurements performed in Pb--Pb collisions at $\sqrt{s_{\rm NN}}$~=~5.02 TeV~\cite{Acharya:2018zuq} (red markers) as a function of centrality (left) and charged-particle density (right)~\cite{Adam:2015ptt, Acharya:2018hhy}. The Pb--Pb points are slightly shifted along the horizontal axis for better visibility in both panels. Bars (boxes) denote statistical (systematic) uncertainties.}
  \label{fig:ptmaxv2_Xe_vs_Pb}
\end{figure}

As mentioned in Sec.~\ref{sec:intro}, the initial state models and transport properties can be further constrained by comparing anisotropic flow coefficients measured in Xe--Xe collisions with those from Pb--Pb collisions. Figures~\ref{fig:v2_Xe_vs_Pb} and~\ref{fig:v3_Xe_vs_Pb} show the \vtwopt{} and \vthreept{} of \pipm{}, \kapm{}, p+\pbar{}, \kzero{}, and \lambdas{} compared with ALICE measurements performed in Pb--Pb collisions at $\snn = 5.02$~TeV~\cite{Acharya:2018zuq} for various centrality intervals. The $v_{\rm n}$ coefficients from Pb--Pb collisions were measured employing the same procedure as described in Sec.~\ref{sec:analysis}, resulting in similar non-flow contributions to $v_{\rm n}$. Ratios of the measurements presented in this paper to a cubic spline fit to the ones performed in Pb--Pb collisions are also given in the figures for each presented centrality interval. The uncertainties in these ratios are obtained by summing the statistical and systematic uncertainties on the Xe--Xe and Pb--Pb measurements in quadrature, and propagating the obtained uncertainties as uncorrelated.

The $v_{\rm n}$ coefficients at low \pt{} are expected to be smaller in Pb--Pb collisions than the corresponding Xe--Xe results due to a larger radial flow in the former, an effect which would be most pronounced in central collisions and for heavier particles. However, the $v_2$ of all particle species in Xe--Xe collisions is systematically above that from Pb--Pb in the entire \pt{} range in the 0--5\% centrality class. The ratios do not depend significantly on \pt{} and particle species within uncertainties, showing $\sim$37\% larger Xe--Xe values. In terms of the initial state, two effects can be responsible for this behaviour. The first relates to the fact that the $^{208}$Pb nucleus is spherical while the $^{129}$Xe nucleus is deformed with parameters of the nuclear-charge density distribution not yet measured directly but extrapolated from neighboring isotopes or predicted (the deformation parameter $\beta_2$ is predicted to be 0.162 in Ref.~\cite{Moller:2015fba} and extrapolated to $0.18 \pm 0.02$ in Ref.~\cite{ALICE-PUBLIC-2018-003}). The second involves initial-state fluctuations being proportional to $A^{-1/2}$~\cite{Bhalerao:2011bp}, where $A$ is the mass number, and the dependence of $\varepsilon_{\rm n}\{2\}$ on the number of sources contributing to it which decreases when the number of sources increases~\cite{Bhalerao:2011bp, Bzdak:2013rya}. These effects imply larger values of $\varepsilon_{2}\{2\}$ for central Xe--Xe collisions than central Pb--Pb collisions, which in turn induce larger $v_2$. However, viscosity is expected to be larger for Xe--Xe collisions as it is proportional to $A^{-1/3}$~\cite{Baier:2007ix} which will decrease $v_2$~\cite{Romatschke:2007mq}. For the 10--20\% centrality interval, the measurements are compatible within uncertainties for the different particle species although a possible suppression of p+\pbar{} $v_2$ from Pb--Pb collisions can be seen for $\pt<1.5$ \GeVc{}. For the 40--50\% centrality class, no differences are observed between the \kzero{} and \lambdas{} \vtwopt{} measured in the two systems within uncertainties, while the $v_2$ of \pipm{}, \kapm{}, and p+\pbar{} from Xe--Xe collisions is $\sim$8\% lower than the corresponding Pb--Pb results. This difference is almost independent of \pt{} within uncertainties although a possible gradual decrease with increasing \pt{} up to 2 \GeVc{} can be seen for p+\pbar{}. The larger $v_2$ values in Pb--Pb collisions might be explained by viscous effects related to the different radial flow and transverse size of the systems since the $\varepsilon_{2}\{2\}$ coefficients are similar in this centrality interval (differences within 1\%)~\cite{Giacalone:2017dud, Eskola:2017bup}. Although $v_3$ is expected to be larger in Xe--Xe compared to Pb--Pb due to larger values of $\varepsilon_{3}\{2\}$ in the same centrality interval~\cite{Giacalone:2017dud, Eskola:2017bup}, the precision of the results does not allow for conclusions to be drawn. The ratios are close to 1 with no significant \pt{} dependence within uncertainties, except for \pipm{} and p+\pbar{} $v_3$ for ${\pt<2}$~\GeVc{} in the 0--10\% centrality class.

The \vtwopt{} of \pipm{}, \kapm{}, and p+\pbar{} measured in Xe--Xe and Pb--Pb collisions is also compared with MUSIC hydrodynamic calculations~\cite{Schenke:2020mbo} in Fig.~\ref{fig:v2_Xe_vs_Pb}. It is worth noting that these calculations employ the same parameters for Xe--Xe and Pb--Pb collisions (see Sec.~\ref{sec:hydroComp}). The Pb--Pb calculations show similar trends to those reported for Xe--Xe collisions: they are in agreement with the measurements for $\pt{}<1$~\GeVc{} and overestimate the data points at higher \pt{}. However, the MUSIC Xe--Xe/Pb--Pb $v_2$ ratios quantitatively reproduce the ones of the measurements up to $\pt = 3$~\GeVc{}. This points to similar differences between the data points and model for both systems. Two potential sources might be responsible for this behavior: improper $\delta f$ corrections, which are introduced in hydrodynamic models to account for non equilibrium processes at freeze-out and are highly model dependent~\cite{Noronha-Hostler:2014dqa}, or sub-optimal tunes of $\eta/s$ and $\zeta/s$.

Figure~\ref{fig:ptmaxv2_Xe_vs_Pb} shows the value of \vtwomax{} of \pipm{} and p+\pbar{} and compares these to the ALICE measurements performed in Pb--Pb collisions at $\snn = 5.02$~TeV~\cite{Acharya:2018zuq} as function of centrality and charged-particle density~\cite{Adam:2015ptt, Acharya:2018hhy}. For all centrality intervals, the \vtwomax{} of p+\pbar{} has similar values in the two collision systems, within uncertainties. The \vtwomax{} of \pipm{} is slightly lower in Xe--Xe collisions in the 5--40\% centrality range. This can be attributed to a different quark density and radial flow at the same centrality in the two systems. Indeed, the \vtwomax{} is the same in Xe--Xe and Pb--Pb collisions for the different particle species within uncertainties when reported as function of charged-particle density.

\section{Summary}
\label{sec:summary}

The elliptic and triangular flow coefficients of \pipm{}, \kapm{}, p+\pbar{}, \kzero{}, and \lambdas{} were measured in Xe--Xe collisions at $\sqrt{s_{\rm NN}}=5.44$~TeV. The magnitude of $v_2$ increases strongly with decreasing centrality up to the 40--50\% centrality interval for all particle species, while $v_3$ shows a weak centrality dependence with a smaller increase than for $v_2$. This indicates that collision geometry dominates the generation of elliptic flow while triangular flow is generated by event-by-event fluctuations in the initial nucleon and gluon densities. For $\pt<3$~\GeVc{}, the $v_{\rm n}$ coefficients show a mass ordering which can be attributed to the interplay between anisotropic flow and radial flow. In this transverse momentum range, MUSIC hydrodynamic calculations reproduce the measured $v_2$ of \pipm{}, \kapm{}, and p+\pbar{} for $\pt{} <1$~\GeVc{}. At intermediate transverse momenta ($3<\pt<$ 8~\GeVc{}), the baryon $v_{\rm n}$ has a magnitude larger than that of mesons, indicating that the particle type dependence persists up to high $\pt$. Furthermore, particles show an approximate grouping by the number of constituent quarks at the level of $\pm 20\%$ for $v_2$. The centrality dependence of the shape evolution of \vtwopt{} is different for \pipm{}, \kapm{}, and p+\pbar{} for $\pt < 2$ \GeVc{}, being more pronounced for p+\pbar{}, but shows no particle type dependence within uncertainties for $\pt \ge 2$ \GeVc{}. Comparing these measurements to those from Pb--Pb collisions at $\snn = 5.02$~TeV, $v_2$ is larger in central collisions at the same centrality and it has smaller value in peripheral collisions.

%%%%%%%%%%%%%%%%%%%%%%%%%%%%%%%%
% end main text 
%%%%%%%%%%%%%%%%%%%%%%%%%%%%%%%%

%%%%% acknowledgements - handled by EB chairs 
\newenvironment{acknowledgement}{\relax}{\relax}
\begin{acknowledgement}
\section*{Acknowledgements}
% add specific acknowledgements here 
% ...but please don't remove the line below: funding agencies
% will be acknowledged with a custom tex file handled by EB chairs after Collab Round
% Version: 2021-07-13

The ALICE Collaboration would like to thank all its engineers and technicians for their invaluable contributions to the construction of the experiment and the CERN accelerator teams for the outstanding performance of the LHC complex.
The ALICE Collaboration gratefully acknowledges the resources and support provided by all Grid centres and the Worldwide LHC Computing Grid (WLCG) collaboration.
The ALICE Collaboration acknowledges the following funding agencies for their support in building and running the ALICE detector:
A. I. Alikhanyan National Science Laboratory (Yerevan Physics Institute) Foundation (ANSL), State Committee of Science and World Federation of Scientists (WFS), Armenia;
Austrian Academy of Sciences, Austrian Science Fund (FWF): [M 2467-N36] and Nationalstiftung f\"{u}r Forschung, Technologie und Entwicklung, Austria;
Ministry of Communications and High Technologies, National Nuclear Research Center, Azerbaijan;
Conselho Nacional de Desenvolvimento Cient\'{\i}fico e Tecnol\'{o}gico (CNPq), Financiadora de Estudos e Projetos (Finep), Funda\c{c}\~{a}o de Amparo \`{a} Pesquisa do Estado de S\~{a}o Paulo (FAPESP) and Universidade Federal do Rio Grande do Sul (UFRGS), Brazil;
Ministry of Education of China (MOEC) , Ministry of Science \& Technology of China (MSTC) and National Natural Science Foundation of China (NSFC), China;
Ministry of Science and Education and Croatian Science Foundation, Croatia;
Centro de Aplicaciones Tecnol\'{o}gicas y Desarrollo Nuclear (CEADEN), Cubaenerg\'{\i}a, Cuba;
Ministry of Education, Youth and Sports of the Czech Republic, Czech Republic;
The Danish Council for Independent Research | Natural Sciences, the VILLUM FONDEN and Danish National Research Foundation (DNRF), Denmark;
Helsinki Institute of Physics (HIP), Finland;
Commissariat \`{a} l'Energie Atomique (CEA) and Institut National de Physique Nucl\'{e}aire et de Physique des Particules (IN2P3) and Centre National de la Recherche Scientifique (CNRS), France;
Bundesministerium f\"{u}r Bildung und Forschung (BMBF) and GSI Helmholtzzentrum f\"{u}r Schwerionenforschung GmbH, Germany;
General Secretariat for Research and Technology, Ministry of Education, Research and Religions, Greece;
National Research, Development and Innovation Office, Hungary;
Department of Atomic Energy Government of India (DAE), Department of Science and Technology, Government of India (DST), University Grants Commission, Government of India (UGC) and Council of Scientific and Industrial Research (CSIR), India;
Indonesian Institute of Science, Indonesia;
Istituto Nazionale di Fisica Nucleare (INFN), Italy;
Institute for Innovative Science and Technology , Nagasaki Institute of Applied Science (IIST), Japanese Ministry of Education, Culture, Sports, Science and Technology (MEXT) and Japan Society for the Promotion of Science (JSPS) KAKENHI, Japan;
Consejo Nacional de Ciencia (CONACYT) y Tecnolog\'{i}a, through Fondo de Cooperaci\'{o}n Internacional en Ciencia y Tecnolog\'{i}a (FONCICYT) and Direcci\'{o}n General de Asuntos del Personal Academico (DGAPA), Mexico;
Nederlandse Organisatie voor Wetenschappelijk Onderzoek (NWO), Netherlands;
The Research Council of Norway, Norway;
Commission on Science and Technology for Sustainable Development in the South (COMSATS), Pakistan;
Pontificia Universidad Cat\'{o}lica del Per\'{u}, Peru;
Ministry of Education and Science, National Science Centre and WUT ID-UB, Poland;
Korea Institute of Science and Technology Information and National Research Foundation of Korea (NRF), Republic of Korea;
Ministry of Education and Scientific Research, Institute of Atomic Physics and Ministry of Research and Innovation and Institute of Atomic Physics, Romania;
Joint Institute for Nuclear Research (JINR), Ministry of Education and Science of the Russian Federation, National Research Centre Kurchatov Institute, Russian Science Foundation and Russian Foundation for Basic Research, Russia;
Ministry of Education, Science, Research and Sport of the Slovak Republic, Slovakia;
National Research Foundation of South Africa, South Africa;
Swedish Research Council (VR) and Knut \& Alice Wallenberg Foundation (KAW), Sweden;
European Organization for Nuclear Research, Switzerland;
Suranaree University of Technology (SUT), National Science and Technology Development Agency (NSDTA) and Office of the Higher Education Commission under NRU project of Thailand, Thailand;
Turkish Energy, Nuclear and Mineral Research Agency (TENMAK), Turkey;
National Academy of  Sciences of Ukraine, Ukraine;
Science and Technology Facilities Council (STFC), United Kingdom;
National Science Foundation of the United States of America (NSF) and United States Department of Energy, Office of Nuclear Physics (DOE NP), United States of America.
\end{acknowledgement}

%%%%%%%% Bibliography 
\bibliographystyle{utphys}   % Remember we use title in the biblio

\bibliography{bibliography}

\providecommand{\href}[2]{#2}\begingroup\raggedright\begin{thebibliography}{10}

\bibitem{Bass:1998vz}
S.~A. Bass, M.~Gyulassy, H.~Stoecker, and W.~Greiner, ``{Signatures of quark
  gluon plasma formation in high-energy heavy ion collisions: A Critical
  review}'', \href{http://dx.doi.org/10.1088/0954-3899/25/3/013}{{\em J. Phys.}
  {\bfseries G25} (1999) R1--R57},
\href{http://arxiv.org/abs/hep-ph/9810281}{{\ttfamily arXiv:hep-ph/9810281
  [hep-ph]}}.
%%CITATION = HEP-PH/9810281;%%.

\bibitem{Ollitrault:1992bk}
J.-Y. Ollitrault, ``{Anisotropy as a signature of transverse collective
  flow}'',
\href{http://dx.doi.org/10.1103/PhysRevD.46.229}{{\em Phys. Rev.} {\bfseries
  D46} (1992) 229--245}.
%%CITATION = PHRVA,D46,229;%%.

\bibitem{Voloshin:1994mz}
S.~Voloshin and Y.~Zhang, ``{Flow study in relativistic nuclear collisions by
  Fourier expansion of Azimuthal particle distributions}'',
  \href{http://dx.doi.org/10.1007/s002880050141}{{\em Z. Phys.} {\bfseries C70}
  (1996) 665--672},
\href{http://arxiv.org/abs/hep-ph/9407282}{{\ttfamily arXiv:hep-ph/9407282
  [hep-ph]}}.
%%CITATION = HEP-PH/9407282;%%.

\bibitem{Poskanzer:1998yz}
A.~M. Poskanzer and S.~A. Voloshin, ``{Methods for analyzing anisotropic flow
  in relativistic nuclear collisions}'',
  \href{http://dx.doi.org/10.1103/PhysRevC.58.1671}{{\em Phys. Rev.} {\bfseries
  C58} (1998) 1671--1678},
\href{http://arxiv.org/abs/nucl-ex/9805001}{{\ttfamily arXiv:nucl-ex/9805001
  [nucl-ex]}}.
%%CITATION = NUCL-EX/9805001;%%.

\bibitem{Bhalerao:2006tp}
R.~S. Bhalerao and J.-Y. Ollitrault, ``{Eccentricity fluctuations and elliptic
  flow at RHIC}'', \href{http://dx.doi.org/10.1016/j.physletb.2006.08.055}{{\em
  Phys. Lett.} {\bfseries B641} (2006) 260--264},
\href{http://arxiv.org/abs/nucl-th/0607009}{{\ttfamily arXiv:nucl-th/0607009
  [nucl-th]}}.
%%CITATION = NUCL-TH/0607009;%%.

\bibitem{Alver:2008zza}
{\bfseries PHOBOS} Collaboration, B.~Alver {\em et~al.}, ``{Importance of
  correlations and fluctuations on the initial source eccentricity in
  high-energy nucleus-nucleus collisions}'',
  \href{http://dx.doi.org/10.1103/PhysRevC.77.014906}{{\em Phys. Rev.}
  {\bfseries C77} (2008) 014906},
\href{http://arxiv.org/abs/0711.3724}{{\ttfamily arXiv:0711.3724 [nucl-ex]}}.
%%CITATION = ARXIV:0711.3724;%%.

\bibitem{Takahashi:2009na}
J.~Takahashi, B.~M. Tavares, W.~L. Qian, R.~Andrade, F.~Grassi, Y.~Hama,
  T.~Kodama, and N.~Xu, ``{Topology studies of hydrodynamics using two particle
  correlation analysis}'',
  \href{http://dx.doi.org/10.1103/PhysRevLett.103.242301}{{\em Phys. Rev.
  Lett.} {\bfseries 103} (2009) 242301},
\href{http://arxiv.org/abs/0902.4870}{{\ttfamily arXiv:0902.4870 [nucl-th]}}.
%%CITATION = ARXIV:0902.4870;%%.

\bibitem{Alver:2010gr}
B.~Alver and G.~Roland, ``{Collision geometry fluctuations and triangular flow
  in heavy-ion collisions}'',
  \href{http://dx.doi.org/10.1103/PhysRevC.82.039903,
  10.1103/PhysRevC.81.054905}{{\em Phys. Rev.} {\bfseries C81} (2010) 054905},
  \href{http://arxiv.org/abs/1003.0194}{{\ttfamily arXiv:1003.0194 [nucl-th]}}.
[Erratum: Phys. Rev.C82,039903(2010)].
%%CITATION = ARXIV:1003.0194;%%.

\bibitem{Alver:2010dn}
B.~H. Alver, C.~Gombeaud, M.~Luzum, and J.-Y. Ollitrault, ``{Triangular flow in
  hydrodynamics and transport theory}'',
  \href{http://dx.doi.org/10.1103/PhysRevC.82.034913}{{\em Phys. Rev.}
  {\bfseries C82} (2010) 034913},
\href{http://arxiv.org/abs/1007.5469}{{\ttfamily arXiv:1007.5469 [nucl-th]}}.
%%CITATION = ARXIV:1007.5469;%%.

\bibitem{Gardim:2011xv}
F.~G. Gardim, F.~Grassi, M.~Luzum, and J.-Y. Ollitrault, ``{Mapping the
  hydrodynamic response to the initial geometry in heavy-ion collisions}'',
  \href{http://dx.doi.org/10.1103/PhysRevC.85.024908}{{\em Phys. Rev.}
  {\bfseries C85} (2012) 024908},
\href{http://arxiv.org/abs/1111.6538}{{\ttfamily arXiv:1111.6538 [nucl-th]}}.
%%CITATION = ARXIV:1111.6538;%%.

\bibitem{Qin:2010pf}
G.-Y. Qin, H.~Petersen, S.~A. Bass, and B.~Muller, ``{Translation of collision
  geometry fluctuations into momentum anisotropies in relativistic heavy-ion
  collisions}'', \href{http://dx.doi.org/10.1103/PhysRevC.82.064903}{{\em Phys.
  Rev. C} {\bfseries 82} (2010) 064903},
  \href{http://arxiv.org/abs/1009.1847}{{\ttfamily arXiv:1009.1847 [nucl-th]}}.

\bibitem{Teaney:2010vd}
D.~Teaney and L.~Yan, ``{Triangularity and Dipole Asymmetry in Heavy Ion
  Collisions}'', \href{http://dx.doi.org/10.1103/PhysRevC.83.064904}{{\em Phys.
  Rev. C} {\bfseries 83} (2011) 064904},
  \href{http://arxiv.org/abs/1010.1876}{{\ttfamily arXiv:1010.1876 [nucl-th]}}.

\bibitem{Arsene:2004fa}
{\bfseries BRAHMS} Collaboration, I.~Arsene {\em et~al.}, ``{Quark gluon plasma
  and color glass condensate at RHIC? The Perspective from the BRAHMS
  experiment}'', \href{http://dx.doi.org/10.1016/j.nuclphysa.2005.02.130}{{\em
  Nucl. Phys.} {\bfseries A757} (2005) 1--27},
\href{http://arxiv.org/abs/nucl-ex/0410020}{{\ttfamily arXiv:nucl-ex/0410020
  [nucl-ex]}}.
%%CITATION = NUCL-EX/0410020;%%.

\bibitem{Adcox:2004mh}
{\bfseries PHENIX} Collaboration, K.~Adcox {\em et~al.}, ``{Formation of dense
  partonic matter in relativistic nucleus-nucleus collisions at RHIC:
  Experimental evaluation by the PHENIX collaboration}'',
  \href{http://dx.doi.org/10.1016/j.nuclphysa.2005.03.086}{{\em Nucl. Phys.}
  {\bfseries A757} (2005) 184--283},
\href{http://arxiv.org/abs/nucl-ex/0410003}{{\ttfamily arXiv:nucl-ex/0410003
  [nucl-ex]}}.
%%CITATION = NUCL-EX/0410003;%%.

\bibitem{Back:2004je}
{\bfseries PHOBOS} Collaboration, B.~B. Back {\em et~al.}, ``{The PHOBOS
  perspective on discoveries at RHIC}'',
  \href{http://dx.doi.org/10.1016/j.nuclphysa.2005.03.084}{{\em Nucl. Phys.}
  {\bfseries A757} (2005) 28--101},
\href{http://arxiv.org/abs/nucl-ex/0410022}{{\ttfamily arXiv:nucl-ex/0410022
  [nucl-ex]}}.
%%CITATION = NUCL-EX/0410022;%%.

\bibitem{Adams:2005dq}
{\bfseries STAR} Collaboration, J.~Adams {\em et~al.}, ``{Experimental and
  theoretical challenges in the search for the quark gluon plasma: The STAR
  Collaboration's critical assessment of the evidence from RHIC collisions}'',
  \href{http://dx.doi.org/10.1016/j.nuclphysa.2005.03.085}{{\em Nucl. Phys.}
  {\bfseries A757} (2005) 102--183},
\href{http://arxiv.org/abs/nucl-ex/0501009}{{\ttfamily arXiv:nucl-ex/0501009
  [nucl-ex]}}.
%%CITATION = NUCL-EX/0501009;%%.

\bibitem{ALICE:2011ab}
{\bfseries ALICE} Collaboration, K.~Aamodt {\em et~al.}, ``{Higher harmonic
  anisotropic flow measurements of charged particles in Pb-Pb collisions at
  $\sqrt{s_{NN}}$=2.76 TeV}'',
  \href{http://dx.doi.org/10.1103/PhysRevLett.107.032301}{{\em Phys. Rev.
  Lett.} {\bfseries 107} (2011) 032301},
\href{http://arxiv.org/abs/1105.3865}{{\ttfamily arXiv:1105.3865 [nucl-ex]}}.
%%CITATION = ARXIV:1105.3865;%%.

\bibitem{Acharya:2018lmh}
{\bfseries ALICE} Collaboration, S.~Acharya {\em et~al.}, ``{Energy dependence
  and fluctuations of anisotropic flow in Pb-Pb collisions at $
  \sqrt{s_{\mathrm{NN}}}=5.02 $ and 2.76 TeV}'',
  \href{http://dx.doi.org/10.1007/JHEP07(2018)103}{{\em JHEP} {\bfseries 07}
  (2018) 103},
\href{http://arxiv.org/abs/1804.02944}{{\ttfamily arXiv:1804.02944 [nucl-ex]}}.
%%CITATION = ARXIV:1804.02944;%%.

\bibitem{ATLAS:2012at}
{\bfseries ATLAS} Collaboration, G.~Aad {\em et~al.}, ``{Measurement of the
  azimuthal anisotropy for charged particle production in $\sqrt{s_{NN}}=2.76$
  TeV lead-lead collisions with the ATLAS detector}'',
  \href{http://dx.doi.org/10.1103/PhysRevC.86.014907}{{\em Phys. Rev.}
  {\bfseries C86} (2012) 014907},
\href{http://arxiv.org/abs/1203.3087}{{\ttfamily arXiv:1203.3087 [hep-ex]}}.
%%CITATION = ARXIV:1203.3087;%%.

\bibitem{Chatrchyan:2013kba}
{\bfseries CMS} Collaboration, S.~Chatrchyan {\em et~al.}, ``{Measurement of
  higher-order harmonic azimuthal anisotropy in PbPb collisions at
  $\sqrt{s_{NN}}$ = 2.76 TeV}'',
  \href{http://dx.doi.org/10.1103/PhysRevC.89.044906}{{\em Phys. Rev.}
  {\bfseries C89} no.~4, (2014) 044906},
\href{http://arxiv.org/abs/1310.8651}{{\ttfamily arXiv:1310.8651 [nucl-ex]}}.
%%CITATION = ARXIV:1310.8651;%%.

\bibitem{Kovtun:2004de}
P.~Kovtun, D.~T. Son, and A.~O. Starinets, ``{Viscosity in strongly interacting
  quantum field theories from black hole physics}'',
  \href{http://dx.doi.org/10.1103/PhysRevLett.94.111601}{{\em Phys. Rev. Lett.}
  {\bfseries 94} (2005) 111601},
\href{http://arxiv.org/abs/hep-th/0405231}{{\ttfamily arXiv:hep-th/0405231
  [hep-th]}}.
%%CITATION = HEP-TH/0405231;%%.

\bibitem{Acharya:2018ihu}
{\bfseries ALICE} Collaboration, S.~Acharya {\em et~al.}, ``{Anisotropic flow
  in Xe-Xe collisions at $\mathbf{\sqrt{s_{\rm{NN}}} = 5.44}$ TeV}'',
  \href{http://dx.doi.org/10.1016/j.physletb.2018.06.059}{{\em Phys. Lett.}
  {\bfseries B784} (2018) 82--95},
\href{http://arxiv.org/abs/1805.01832}{{\ttfamily arXiv:1805.01832 [nucl-ex]}}.
%%CITATION = ARXIV:1805.01832;%%.

\bibitem{Sirunyan:2019wqp}
{\bfseries CMS} Collaboration, A.~M. Sirunyan {\em et~al.}, ``{Charged-particle
  angular correlations in XeXe collisions at $\sqrt{s_{_\mathrm{NN}}}=$ 5.44
  TeV}'', \href{http://dx.doi.org/10.1103/PhysRevC.100.044902}{{\em Phys. Rev.
  C} {\bfseries 100} no.~4, (2019) 044902},
  \href{http://arxiv.org/abs/1901.07997}{{\ttfamily arXiv:1901.07997
  [hep-ex]}}.

\bibitem{Aad:2019xmh}
{\bfseries ATLAS} Collaboration, G.~Aad {\em et~al.}, ``{Measurement of the
  azimuthal anisotropy of charged-particle production in $Xe+Xe$ collisions at
  $\sqrt{s_{\mathrm{NN}}}=5.44$ TeV with the ATLAS detector}'',
  \href{http://dx.doi.org/10.1103/PhysRevC.101.024906}{{\em Phys. Rev. C}
  {\bfseries 101} no.~2, (2020) 024906},
  \href{http://arxiv.org/abs/1911.04812}{{\ttfamily arXiv:1911.04812
  [nucl-ex]}}.

\bibitem{Giacalone:2017dud}
G.~Giacalone, J.~Noronha-Hostler, M.~Luzum, and J.-Y. Ollitrault,
  ``{Hydrodynamic predictions for 5.44 TeV Xe+Xe collisions}'',
  \href{http://dx.doi.org/10.1103/PhysRevC.97.034904}{{\em Phys. Rev.}
  {\bfseries C97} no.~3, (2018) 034904},
\href{http://arxiv.org/abs/1711.08499}{{\ttfamily arXiv:1711.08499 [nucl-th]}}.
%%CITATION = ARXIV:1711.08499;%%.

\bibitem{Eskola:2017bup}
K.~J. Eskola, H.~Niemi, R.~Paatelainen, and K.~Tuominen, ``{Predictions for
  multiplicities and flow harmonics in 5.44 TeV Xe+Xe collisions at the CERN
  Large Hadron Collider}'',
  \href{http://dx.doi.org/10.1103/PhysRevC.97.034911}{{\em Phys. Rev.}
  {\bfseries C97} no.~3, (2018) 034911},
\href{http://arxiv.org/abs/1711.09803}{{\ttfamily arXiv:1711.09803 [hep-ph]}}.
%%CITATION = ARXIV:1711.09803;%%.

\bibitem{Qiu:2011iv}
Z.~Qiu and U.~W. Heinz, ``{Event-by-event shape and flow fluctuations of
  relativistic heavy-ion collision fireballs}'',
  \href{http://dx.doi.org/10.1103/PhysRevC.84.024911}{{\em Phys. Rev.}
  {\bfseries C84} (2011) 024911},
\href{http://arxiv.org/abs/1104.0650}{{\ttfamily arXiv:1104.0650 [nucl-th]}}.
%%CITATION = ARXIV:1104.0650;%%.

\bibitem{Huovinen:2001cy}
P.~Huovinen, P.~Kolb, U.~W. Heinz, P.~Ruuskanen, and S.~Voloshin, ``{Radial and
  elliptic flow at RHIC: Further predictions}'',
  \href{http://dx.doi.org/10.1016/S0370-2693(01)00219-2}{{\em Phys. Lett. B}
  {\bfseries 503} (2001) 58--64},
  \href{http://arxiv.org/abs/hep-ph/0101136}{{\ttfamily arXiv:hep-ph/0101136}}.

\bibitem{Shen:2011eg}
C.~Shen, U.~Heinz, P.~Huovinen, and H.~Song, ``{Radial and elliptic flow in
  Pb+Pb collisions at the Large Hadron Collider from viscous hydrodynamic}'',
  \href{http://dx.doi.org/10.1103/PhysRevC.84.044903}{{\em Phys. Rev. C}
  {\bfseries 84} (2011) 044903},
  \href{http://arxiv.org/abs/1105.3226}{{\ttfamily arXiv:1105.3226 [nucl-th]}}.

\bibitem{Abelev:2014pua}
{\bfseries ALICE} Collaboration, B.~Abelev {\em et~al.}, ``{Elliptic flow of
  identified hadrons in Pb-Pb collisions at $ \sqrt{s_{\mathrm{NN}}}=2.76 $
  TeV}'', \href{http://dx.doi.org/10.1007/JHEP06(2015)190}{{\em JHEP}
  {\bfseries 06} (2015) 190}, \href{http://arxiv.org/abs/1405.4632}{{\ttfamily
  arXiv:1405.4632 [nucl-ex]}}.

\bibitem{Adam:2016nfo}
{\bfseries ALICE} Collaboration, J.~Adam {\em et~al.}, ``{Higher harmonic flow
  coefficients of identified hadrons in Pb-Pb collisions at $\sqrt{s_{\rm NN}}$
  = 2.76 TeV}'', \href{http://dx.doi.org/10.1007/JHEP09(2016)164}{{\em JHEP}
  {\bfseries 09} (2016) 164}, \href{http://arxiv.org/abs/1606.06057}{{\ttfamily
  arXiv:1606.06057 [nucl-ex]}}.

\bibitem{Acharya:2018zuq}
{\bfseries ALICE} Collaboration, S.~Acharya {\em et~al.}, ``{Anisotropic flow
  of identified particles in Pb-Pb collisions at $
  {\sqrt{s}}_{\mathrm{NN}}=5.02 $ TeV}'',
  \href{http://dx.doi.org/10.1007/JHEP09(2018)006}{{\em JHEP} {\bfseries 09}
  (2018) 006},
\href{http://arxiv.org/abs/1805.04390}{{\ttfamily arXiv:1805.04390 [nucl-ex]}}.
%%CITATION = ARXIV:1805.04390;%%.

\bibitem{Adams:2003am}
{\bfseries STAR} Collaboration, J.~Adams {\em et~al.}, ``{Particle type
  dependence of azimuthal anisotropy and nuclear modification of particle
  production in Au + Au collisions at $\sqrt{s_{\rm NN}}=200$ GeV}'',
  \href{http://dx.doi.org/10.1103/PhysRevLett.92.052302}{{\em Phys. Rev. Lett.}
  {\bfseries 92} (2004) 052302},
  \href{http://arxiv.org/abs/nucl-ex/0306007}{{\ttfamily
  arXiv:nucl-ex/0306007}}.

\bibitem{Adler:2003kt}
{\bfseries PHENIX} Collaboration, S.~Adler {\em et~al.}, ``{Elliptic flow of
  identified hadrons in Au+Au collisions at $\sqrt{s_{\rm NN}}=200$ GeV}'',
  \href{http://dx.doi.org/10.1103/PhysRevLett.91.182301}{{\em Phys. Rev. Lett.}
  {\bfseries 91} (2003) 182301},
  \href{http://arxiv.org/abs/nucl-ex/0305013}{{\ttfamily
  arXiv:nucl-ex/0305013}}.

\bibitem{Abelev:2012di}
{\bfseries ALICE} Collaboration, B.~Abelev {\em et~al.}, ``{Anisotropic flow of
  charged hadrons, pions and (anti-)protons measured at high transverse
  momentum in Pb-Pb collisions at $\sqrt{s_{NN}}$=2.76 TeV}'',
  \href{http://dx.doi.org/10.1016/j.physletb.2012.12.066}{{\em Phys. Lett. B}
  {\bfseries 719} (2013) 18--28},
  \href{http://arxiv.org/abs/1205.5761}{{\ttfamily arXiv:1205.5761 [nucl-ex]}}.

\bibitem{Abelev:2007qg}
{\bfseries STAR} Collaboration, B.~Abelev {\em et~al.}, ``{Mass, quark-number,
  and $\sqrt{s_{NN}}$ dependence of the second and fourth flow harmonics in
  ultra-relativistic nucleus-nucleus collisions}'',
  \href{http://dx.doi.org/10.1103/PhysRevC.75.054906}{{\em Phys. Rev. C}
  {\bfseries 75} (2007) 054906},
  \href{http://arxiv.org/abs/nucl-ex/0701010}{{\ttfamily
  arXiv:nucl-ex/0701010}}.

\bibitem{Adare:2012vq}
{\bfseries PHENIX} Collaboration, A.~Adare {\em et~al.}, ``{Deviation from
  quark-number scaling of the anisotropy parameter $v_2$ of pions, kaons, and
  protons in Au+Au collisions at $\sqrt{s_{NN}} = 200$ GeV}'',
  \href{http://dx.doi.org/10.1103/PhysRevC.85.064914}{{\em Phys. Rev. C}
  {\bfseries 85} (2012) 064914},
  \href{http://arxiv.org/abs/1203.2644}{{\ttfamily arXiv:1203.2644 [nucl-ex]}}.

\bibitem{Molnar:2003ff}
D.~Molnar and S.~A. Voloshin, ``{Elliptic flow at large transverse momenta from
  quark coalescence}'',
  \href{http://dx.doi.org/10.1103/PhysRevLett.91.092301}{{\em Phys. Rev. Lett.}
  {\bfseries 91} (2003) 092301},
  \href{http://arxiv.org/abs/nucl-th/0302014}{{\ttfamily
  arXiv:nucl-th/0302014}}.

\bibitem{Greco:2003mm}
V.~Greco, C.~Ko, and P.~Levai, ``{Parton coalescence at RHIC}'',
  \href{http://dx.doi.org/10.1103/PhysRevC.68.034904}{{\em Phys. Rev. C}
  {\bfseries 68} (2003) 034904},
  \href{http://arxiv.org/abs/nucl-th/0305024}{{\ttfamily
  arXiv:nucl-th/0305024}}.

\bibitem{Fries:2003kq}
R.~Fries, B.~Muller, C.~Nonaka, and S.~Bass, ``{Hadron production in heavy ion
  collisions: Fragmentation and recombination from a dense parton phase}'',
  \href{http://dx.doi.org/10.1103/PhysRevC.68.044902}{{\em Phys. Rev. C}
  {\bfseries 68} (2003) 044902},
  \href{http://arxiv.org/abs/nucl-th/0306027}{{\ttfamily
  arXiv:nucl-th/0306027}}.

\bibitem{Werner:2012xh}
K.~Werner, I.~Karpenko, M.~Bleicher, T.~Pierog, and S.~Porteboeuf-Houssais,
  ``{Jets, Bulk Matter, and their Interaction in Heavy Ion Collisions at
  Several TeV}'', \href{http://dx.doi.org/10.1103/PhysRevC.85.064907}{{\em
  Phys. Rev. C} {\bfseries 85} (2012) 064907},
  \href{http://arxiv.org/abs/1203.5704}{{\ttfamily arXiv:1203.5704 [nucl-th]}}.

\bibitem{Sato:1981ez}
H.~Sato and K.~Yazaki, ``{On the coalescence model for high-energy nuclear
  reactions}'', \href{http://dx.doi.org/10.1016/0370-2693(81)90976-X}{{\em
  Phys. Lett. B} {\bfseries 98} (1981) 153--157}.

\bibitem{Dover:1991zn}
C.~B. Dover, U.~W. Heinz, E.~Schnedermann, and J.~Zimanyi, ``{Covariant
  coalescence model for relativistically expanding systems}'',
  \href{http://dx.doi.org/10.1103/PhysRevC.44.1636}{{\em Phys. Rev. C}
  {\bfseries 44} (1991) 1636--1654}.

\bibitem{Adler:2002pu}
{\bfseries STAR} Collaboration, C.~Adler {\em et~al.}, ``{Elliptic flow from
  two and four particle correlations in Au+Au collisions at $\sqrt{s_{\rm
  NN}}=130$ GeV}'', \href{http://dx.doi.org/10.1103/PhysRevC.66.034904}{{\em
  Phys. Rev. C} {\bfseries 66} (2002) 034904},
  \href{http://arxiv.org/abs/nucl-ex/0206001}{{\ttfamily
  arXiv:nucl-ex/0206001}}.

\bibitem{Voloshin:2008dg}
S.~A. Voloshin, A.~M. Poskanzer, and R.~Snellings, ``{Collective phenomena in
  non-central nuclear collisions}'',
  \href{http://dx.doi.org/10.1007/978-3-642-01539-7_10}{{\em Landolt-Bornstein}
  {\bfseries 23} (2010) 293--333},
  \href{http://arxiv.org/abs/0809.2949}{{\ttfamily arXiv:0809.2949 [nucl-ex]}}.

\bibitem{Luzum:2012da}
M.~Luzum and J.-Y. Ollitrault, ``{Eliminating experimental bias in
  anisotropic-flow measurements of high-energy nuclear collisions}'',
  \href{http://dx.doi.org/10.1103/PhysRevC.87.044907}{{\em Phys. Rev. C}
  {\bfseries 87} no.~4, (2013) 044907},
  \href{http://arxiv.org/abs/1209.2323}{{\ttfamily arXiv:1209.2323 [nucl-ex]}}.

\bibitem{Aamodt:2008zz}
{\bfseries ALICE} Collaboration, K.~Aamodt {\em et~al.}, ``{The ALICE
  experiment at the CERN LHC}'',
  \href{http://dx.doi.org/10.1088/1748-0221/3/08/S08002}{{\em JINST} {\bfseries
  3} (2008) S08002}.

\bibitem{Abelev:2014ffa}
{\bfseries ALICE} Collaboration, B.~Abelev {\em et~al.}, ``{Performance of the
  ALICE Experiment at the CERN LHC}'',
  \href{http://dx.doi.org/10.1142/S0217751X14300440}{{\em Int. J. Mod. Phys. A}
  {\bfseries 29} (2014) 1430044},
  \href{http://arxiv.org/abs/1402.4476}{{\ttfamily arXiv:1402.4476 [nucl-ex]}}.

\bibitem{Aamodt:2010aa}
{\bfseries ALICE} Collaboration, K.~Aamodt {\em et~al.}, ``{Alignment of the
  ALICE Inner Tracking System with cosmic-ray tracks}'',
  \href{http://dx.doi.org/10.1088/1748-0221/5/03/P03003}{{\em JINST} {\bfseries
  5} (2010) P03003}, \href{http://arxiv.org/abs/1001.0502}{{\ttfamily
  arXiv:1001.0502 [physics.ins-det]}}.

\bibitem{Alme:2010ke}
J.~Alme {\em et~al.}, ``{The ALICE TPC, a large 3-dimensional tracking device
  with fast readout for ultra-high multiplicity events}'',
  \href{http://dx.doi.org/10.1016/j.nima.2010.04.042}{{\em Nucl. Instrum. Meth.
  A} {\bfseries 622} (2010) 316--367},
  \href{http://arxiv.org/abs/1001.1950}{{\ttfamily arXiv:1001.1950
  [physics.ins-det]}}.

\bibitem{Akindinov:2013tea}
A.~Akindinov {\em et~al.}, ``{Performance of the ALICE Time-Of-Flight detector
  at the LHC}'', \href{http://dx.doi.org/10.1140/epjp/i2013-13044-x}{{\em Eur.
  Phys. J. Plus} {\bfseries 128} (2013) 44}.

\bibitem{Abbas:2013taa}
{\bfseries ALICE} Collaboration, E.~Abbas {\em et~al.}, ``{Performance of the
  ALICE VZERO system}'',
  \href{http://dx.doi.org/10.1088/1748-0221/8/10/P10016}{{\em JINST} {\bfseries
  8} (2013) P10016}, \href{http://arxiv.org/abs/1306.3130}{{\ttfamily
  arXiv:1306.3130 [nucl-ex]}}.

\bibitem{Bondila:2005xy}
M.~Bondila {\em et~al.}, ``{ALICE T0 detector}'',
  \href{http://dx.doi.org/10.1109/TNS.2005.856900}{{\em IEEE Trans. Nucl. Sci.}
  {\bfseries 52} (2005) 1705--1711}.

\bibitem{ALICE-PUBLIC-2018-003}
{\bfseries ALICE} Collaboration, S.~Acharya {\em et~al.}, ``{Centrality
  determination using the Glauber model in Xe-Xe collisions at $\sqrt{s_{\rm
  NN}} = 5.44$ TeV}'', {\em ALICE-PUBLIC-2018-003} (2018) 1--23.
  \url{{http://cds.cern.ch/record/2315401}}.

\bibitem{Arnaldi:1999zz}
R.~Arnaldi {\em et~al.}, ``{The Zero degree calorimeters for the ALICE
  experiment}'', \href{http://dx.doi.org/10.1016/j.nima.2008.04.009}{{\em Nucl.
  Instrum. Meth. A} {\bfseries 581} (2007) 397--401}. [Erratum:
  Nucl.Instrum.Meth.A 604, 765 (2009)].

\bibitem{Acharya:2018eaq}
{\bfseries ALICE} Collaboration, S.~Acharya {\em et~al.}, ``{Transverse
  momentum spectra and nuclear modification factors of charged particles in
  Xe-Xe collisions at $\sqrt{s_{\rm NN}}$ = 5.44 TeV}'',
  \href{http://dx.doi.org/10.1016/j.physletb.2018.10.052}{{\em Phys. Lett. B}
  {\bfseries 788} (2019) 166--179},
  \href{http://arxiv.org/abs/1805.04399}{{\ttfamily arXiv:1805.04399
  [nucl-ex]}}.

\bibitem{Abelev:2013vea}
{\bfseries ALICE} Collaboration, B.~Abelev {\em et~al.}, ``{Centrality
  dependence of $\pi$, K, p production in Pb-Pb collisions at $\sqrt{s_{NN}}$ =
  2.76 TeV}'', \href{http://dx.doi.org/10.1103/PhysRevC.88.044910}{{\em Phys.
  Rev. C} {\bfseries 88} (2013) 044910},
  \href{http://arxiv.org/abs/1303.0737}{{\ttfamily arXiv:1303.0737 [hep-ex]}}.

\bibitem{armpod}
J.~Podolanski and R.~Armenteros, ``Iii. analysis of v-events'',
  \href{http://dx.doi.org/10.1080/14786440108520416}{{\em The London,
  Edinburgh, and Dublin Philosophical Magazine and Journal of Science}
  {\bfseries 45} no.~360, (1954) 13--30}.

\bibitem{Selyuzhenkov:2007zi}
I.~Selyuzhenkov and S.~Voloshin, ``{Effects of non-uniform acceptance in
  anisotropic flow measurement}'',
  \href{http://dx.doi.org/10.1103/PhysRevC.77.034904}{{\em Phys. Rev. C}
  {\bfseries 77} (2008) 034904},
  \href{http://arxiv.org/abs/0707.4672}{{\ttfamily arXiv:0707.4672 [nucl-th]}}.

\bibitem{Borghini:2004ra}
N.~Borghini and J.~Ollitrault, ``{Azimuthally sensitive correlations in
  nucleus-nucleus collisions}'',
  \href{http://dx.doi.org/10.1103/PhysRevC.70.064905}{{\em Phys. Rev. C}
  {\bfseries 70} (2004) 064905},
  \href{http://arxiv.org/abs/nucl-th/0407041}{{\ttfamily
  arXiv:nucl-th/0407041}}.

\bibitem{Barlow:2002yb}
R.~Barlow, ``{Systematic errors: Facts and fictions}'', in {\em {Conference on
  Advanced Statistical Techniques in Particle Physics}}, pp.~134--144.
\newblock 7, 2002.
\newblock \href{http://arxiv.org/abs/hep-ex/0207026}{{\ttfamily
  arXiv:hep-ex/0207026}}.

\bibitem{Song:2007fn}
H.~Song and U.~W. Heinz, ``{Suppression of elliptic flow in a minimally viscous
  quark-gluon plasma}'',
  \href{http://dx.doi.org/10.1016/j.physletb.2007.11.019}{{\em Phys. Lett. B}
  {\bfseries 658} (2008) 279--283},
  \href{http://arxiv.org/abs/0709.0742}{{\ttfamily arXiv:0709.0742 [nucl-th]}}.

\bibitem{Song:2013qma}
H.~Song, S.~Bass, and U.~W. Heinz, ``{Spectra and elliptic flow for identified
  hadrons in 2.76A TeV Pb + Pb collisions}'',
  \href{http://dx.doi.org/10.1103/PhysRevC.89.034919}{{\em Phys. Rev. C}
  {\bfseries 89} no.~3, (2014) 034919},
  \href{http://arxiv.org/abs/1311.0157}{{\ttfamily arXiv:1311.0157 [nucl-th]}}.

\bibitem{Acharya:2021ljw}
{\bfseries ALICE} Collaboration, S.~Acharya {\em et~al.}, ``{Production of
  pions, kaons, (anti-)protons and $\phi$ mesons in Xe-Xe collisions at
  $\sqrt{s_{\rm NN}} = 5.44$ TeV}'',
  \href{http://arxiv.org/abs/2101.03100}{{\ttfamily arXiv:2101.03100
  [nucl-ex]}}.

\bibitem{Fries:2008hs}
R.~J. Fries, V.~Greco, and P.~Sorensen, ``{Coalescence Models For Hadron
  Formation From Quark Gluon Plasma}'',
  \href{http://dx.doi.org/10.1146/annurev.nucl.58.110707.171134}{{\em Ann. Rev.
  Nucl. Part. Sci.} {\bfseries 58} (2008) 177--205},
  \href{http://arxiv.org/abs/0807.4939}{{\ttfamily arXiv:0807.4939 [nucl-th]}}.

\bibitem{Borghini:2005kd}
N.~Borghini and J.-Y. Ollitrault, ``{Momentum spectra, anisotropic flow, and
  ideal fluids}'', \href{http://dx.doi.org/10.1016/j.physletb.2006.09.062}{{\em
  Phys. Lett.} {\bfseries B642} (2006) 227--231},
\href{http://arxiv.org/abs/nucl-th/0506045}{{\ttfamily arXiv:nucl-th/0506045
  [nucl-th]}}.
%%CITATION = NUCL-TH/0506045;%%.

\bibitem{Schenke:2020mbo}
B.~Schenke, C.~Shen, and P.~Tribedy, ``{Running the gamut of high energy
  nuclear collisions}'',
  \href{http://dx.doi.org/10.1103/PhysRevC.102.044905}{{\em Phys. Rev. C}
  {\bfseries 102} no.~4, (2020) 044905},
  \href{http://arxiv.org/abs/2005.14682}{{\ttfamily arXiv:2005.14682
  [nucl-th]}}.

\bibitem{Schenke:2010rr}
B.~Schenke, S.~Jeon, and C.~Gale, ``{Elliptic and triangular flow in
  event-by-event (3+1)D viscous hydrodynamics}'',
  \href{http://dx.doi.org/10.1103/PhysRevLett.106.042301}{{\em Phys. Rev.
  Lett.} {\bfseries 106} (2011) 042301},
\href{http://arxiv.org/abs/1009.3244}{{\ttfamily arXiv:1009.3244 [hep-ph]}}.
%%CITATION = ARXIV:1009.3244;%%.

\bibitem{glasma1}
B.~Schenke, P.~Tribedy, and R.~Venugopalan, ``Fluctuating glasma initial
  conditions and flow in heavy ion collisions'',
  \href{http://dx.doi.org/10.1103/PhysRevLett.108.252301}{{\em Phys. Rev.
  Lett.} {\bfseries 108} (Jun, 2012) 252301}.

\bibitem{glasma2}
B.~Schenke, P.~Tribedy, and R.~Venugopalan, ``Event-by-event gluon
  multiplicity, energy density, and eccentricities in ultrarelativistic
  heavy-ion collisions'',
  \href{http://dx.doi.org/10.1103/PhysRevC.86.034908}{{\em Phys. Rev. C}
  {\bfseries 86} (Sep, 2012) 034908}.

\bibitem{glasmacascade}
S.~A. Bass {\em et~al.}, ``{Microscopic models for ultrarelativistic heavy ion
  collisions}'', \href{http://dx.doi.org/10.1016/S0146-6410(98)00058-1}{{\em
  Prog. Part. Nucl. Phys.} {\bfseries 41} (1998) 255--369},
  \href{http://arxiv.org/abs/nucl-th/9803035}{{\ttfamily arXiv:nucl-th/9803035
  [nucl-th]}}.
[Prog. Part. Nucl. Phys.41,225(1998)].
%%CITATION = NUCL-TH/9803035;%%.

\bibitem{bleichercascade}
M.~Bleicher {\em et~al.}, ``{Relativistic hadron hadron collisions in the
  ultrarelativistic quantum molecular dynamics model}'',
  \href{http://dx.doi.org/10.1088/0954-3899/25/9/308}{{\em J. Phys.} {\bfseries
  G25} (1999) 1859--1896},
\href{http://arxiv.org/abs/hep-ph/9909407}{{\ttfamily arXiv:hep-ph/9909407
  [hep-ph]}}.
%%CITATION = HEP-PH/9909407;%%.

\bibitem{Adam:2015ptt}
{\bfseries ALICE} Collaboration, J.~Adam {\em et~al.}, ``{Centrality dependence
  of the charged-particle multiplicity density at midrapidity in Pb-Pb
  collisions at $\sqrt{s_{\rm NN}}$ = 5.02 TeV}'',
  \href{http://dx.doi.org/10.1103/PhysRevLett.116.222302}{{\em Phys. Rev.
  Lett.} {\bfseries 116} no.~22, (2016) 222302},
  \href{http://arxiv.org/abs/1512.06104}{{\ttfamily arXiv:1512.06104
  [nucl-ex]}}.

\bibitem{Acharya:2018hhy}
{\bfseries ALICE} Collaboration, S.~Acharya {\em et~al.}, ``{Centrality and
  pseudorapidity dependence of the charged-particle multiplicity density in
  Xe\textendash{}Xe collisions at $\sqrt{s_{\rm NN}}$ = 5.44 TeV}'',
  \href{http://dx.doi.org/10.1016/j.physletb.2018.12.048}{{\em Phys. Lett. B}
  {\bfseries 790} (2019) 35--48},
  \href{http://arxiv.org/abs/1805.04432}{{\ttfamily arXiv:1805.04432
  [nucl-ex]}}.

\bibitem{Moller:2015fba}
P.~M\"oller, A.~Sierk, T.~Ichikawa, and H.~Sagawa, ``{Nuclear ground-state
  masses and deformations: FRDM(2012)}'',
  \href{http://dx.doi.org/10.1016/j.adt.2015.10.002}{{\em Atom. Data Nucl. Data
  Tabl.} {\bfseries 109-110} (2016) 1--204},
  \href{http://arxiv.org/abs/1508.06294}{{\ttfamily arXiv:1508.06294
  [nucl-th]}}.

\bibitem{Bhalerao:2011bp}
R.~S. Bhalerao, M.~Luzum, and J.-Y. Ollitrault, ``{Understanding anisotropy
  generated by fluctuations in heavy-ion collisions}'',
  \href{http://dx.doi.org/10.1103/PhysRevC.84.054901}{{\em Phys. Rev. C}
  {\bfseries 84} (2011) 054901},
  \href{http://arxiv.org/abs/1107.5485}{{\ttfamily arXiv:1107.5485 [nucl-th]}}.

\bibitem{Bzdak:2013rya}
A.~Bzdak, P.~Bozek, and L.~McLerran, ``{Fluctuation induced equality of
  multi-particle eccentricities for four or more particles}'',
  \href{http://dx.doi.org/10.1016/j.nuclphysa.2014.03.007}{{\em Nucl. Phys. A}
  {\bfseries 927} (2014) 15--23},
  \href{http://arxiv.org/abs/1311.7325}{{\ttfamily arXiv:1311.7325 [hep-ph]}}.

\bibitem{Baier:2007ix}
R.~Baier, P.~Romatschke, D.~T. Son, A.~O. Starinets, and M.~A. Stephanov,
  ``{Relativistic viscous hydrodynamics, conformal invariance, and
  holography}'', \href{http://dx.doi.org/10.1088/1126-6708/2008/04/100}{{\em
  JHEP} {\bfseries 04} (2008) 100},
  \href{http://arxiv.org/abs/0712.2451}{{\ttfamily arXiv:0712.2451 [hep-th]}}.

\bibitem{Romatschke:2007mq}
P.~Romatschke and U.~Romatschke, ``{Viscosity Information from Relativistic
  Nuclear Collisions: How Perfect is the Fluid Observed at RHIC?}'',
  \href{http://dx.doi.org/10.1103/PhysRevLett.99.172301}{{\em Phys. Rev. Lett.}
  {\bfseries 99} (2007) 172301},
  \href{http://arxiv.org/abs/0706.1522}{{\ttfamily arXiv:0706.1522 [nucl-th]}}.

\bibitem{Noronha-Hostler:2014dqa}
J.~Noronha-Hostler, J.~Noronha, and F.~Grassi, ``{Bulk viscosity-driven
  suppression of shear viscosity effects on the flow harmonics at energies
  available at the BNL Relativistic Heavy Ion Collider}'',
  \href{http://dx.doi.org/10.1103/PhysRevC.90.034907}{{\em Phys. Rev. C}
  {\bfseries 90} no.~3, (2014) 034907},
  \href{http://arxiv.org/abs/1406.3333}{{\ttfamily arXiv:1406.3333 [nucl-th]}}.

\end{thebibliography}\endgroup
%\input {bibliography.tex}  

%%%%%%%%%%%%%%%%%%%%%%%%%%%%%%%%
% Appendices: yours (if any) + authorlist
%%%%%%%%%%%%%%%%%%%%%%%%%%%%%%%%
\newpage
\appendix

%
%\input{} % put your appendices here (if any)
%

%%%%% Authorlist - please do not touch: handled by EB chairs 
\section{The ALICE Collaboration}
\label{app:collab}

\begin{flushleft} 

S.~Acharya$^{\rm 141}$, 
D.~Adamov\'{a}$^{\rm 96}$, 
A.~Adler$^{\rm 74}$, 
G.~Aglieri Rinella$^{\rm 34}$, 
M.~Agnello$^{\rm 30}$, 
N.~Agrawal$^{\rm 54}$, 
Z.~Ahammed$^{\rm 141}$, 
S.~Ahmad$^{\rm 16}$, 
S.U.~Ahn$^{\rm 76}$, 
I.~Ahuja$^{\rm 38}$
% ...

\bigskip

\bigskip 

\textbf{\Large Affiliation Notes}

\bigskip 

% ...
$^{\rm VII}$ Also at: University of Kansas, Lawrence, Kansas, United States\\

\bigskip

\bigskip 

\textbf{\Large Collaboration Institutes}

\bigskip 

% ...
$^{16}$ Department of Physics, Aligarh Muslim University, Aligarh, India\\
% ...
$^{30}$ Dipartimento DISAT del Politecnico and Sezione INFN, Turin, Italy\\
% ...
$^{34}$ European Organization for Nuclear Research (CERN), Geneva, Switzerland\\
% ...
$^{38}$ Faculty of Science, P.J.~\v{S}af\'{a}rik University, Ko\v{s}ice, Slovakia\\
% ...
$^{54}$ INFN, Sezione di Bologna, Bologna, Italy\\
% ...
$^{74}$ Johann-Wolfgang-Goethe Universit\"{a}t Frankfurt Institut f\"{u}r Informatik, Fachbereich Informatik und Mathematik, Frankfurt, Germany\\
% ...
$^{76}$ Korea Institute of Science and Technology Information, Daejeon, Republic of Korea\\
% ...
$^{96}$ Nuclear Physics Institute of the Czech Academy of Sciences, \v{R}e\v{z} u Prahy, Czech Republic\\
% ...
$^{141}$ Variable Energy Cyclotron Centre, Homi Bhabha National Institute, Kolkata, India\\
% ...

\bigskip 

\end{flushleft}   
\end{document}